\documentclass[preprint,11pt]{article}
\usepackage{epsfig}
\usepackage{amssymb,amsfonts,amsmath,amsthm,amscd}
\usepackage{graphicx}

\newtheorem{propo}{Proposition}
\newtheorem{lemma}{Lemma}
\newtheorem{thm}{Theorem}

\def\prooft{\hspace{0.5cm}{\bf Proof:}\hspace{0.1cm}}
\def\endproof{\hfill$\Box$\vspace{0.4cm}}

\def\tC{\widetilde{C}}

\def\fp{\widehat{f}}
\def\ps{\mu_*}

\def\cX{{\cal X}}
\def\<{\langle}
\def\>{\rangle}

\def\prob{{\mathbb P}}
\def\E{{\mathbb E}}
\def\ve{\varepsilon}
\def\Ball{{\sf B}}
\def\sTV{{\mbox{\tiny\rm TV}}}

\def\pd{\partial}
\def\de{{\rm d}}

\def\reals{{\mathbb R}}
\def\ind{{\mathbb I}}

\def\l|{{\big|\big|}}
\def\r|{{\big|\big|}}
\def\sign{{\rm sign}}

\def\Ev{{\cal A}}
\def\NEv{\overline{\cal A}}

\def\coeff{{\sf coeff}}

\def\Ep{{\sf E}}
\def\Pp{{\sf P}}

\def\T{{\sf T}}
\def\orh{\overline{\rho}}
\def\ph{\widehat{\phi}}

\def\cU{{\cal U}}
\def\Tree{{\sf T}}

\begin{document}

\date{\today}

\title{Rigorous Inequalities between Length and Time Scales in
Glassy Systems}

\author{
   Andrea Montanari\footnote{Laboratoire de Physique Th\'{e}orique
 de l'Ecole Normale Sup\'{e}rieure
(UMR 8549, Unit{\'e}   Mixte de Recherche du
CNRS  et de l' ENS).} \ \ and  \ \    
  Guilhem Semerjian\footnote{ Dipartimento di Fisica, 
CNR-INFM (UdR and SMC centre),
Universit\`{a} di Roma ``La Sapienza''.}
}

\maketitle

\begin{abstract}
Glassy systems are characterized by an extremely sluggish dynamics without
any simple sign of long range order. It is a debated question whether 
a correct description of such phenomenon requires the emergence of a 
large correlation length. We prove rigorous bounds between length and time
scales implying the growth of a properly defined length when the relaxation
time increases. Our results are valid in a rather general setting, which covers
finite-dimensional and mean field systems. 

As an illustration, we discuss the Glauber (heat bath) dynamics
of $p$-spin glass models on random regular graphs. We present the
first proof that a model of this type undergoes a purely dynamical 
phase transition not accompanied by any thermodynamic singularity.
\end{abstract}
%
%
\section{Introduction}

A broad class of liquids show a dramatic slowing down of their dynamics
as the temperature is lowered (or the density increased) without 
any simple sign of long range order setting in
\cite{LesHouches}. In particular, the static structure factor 
$S(\vec{k}) = \<\rho_{\vec k}\rho_{-\vec k}\>$ is hardly distinguishable
from the one of a liquid, even if the system has become a solid from a 
dynamical point of view.
We shall generically refer to systems displaying analogous 
phenomena as `glassy'.

The above features are reproduced in mode-coupling theory, as
well as in some mean-field models, where the slowing down is promoted
to a real {\em dynamical phase transition} without any simple
{\em static} signature \cite{DynamicsReview}. 
In both these theories, only short range correlations 
are taken into account, thus leading to the idea that they are the only 
responsible for the slow dynamics.

Physical commonsense suggests however that, in systems with finite range 
interactions, only cooperative effects on large length scales may lead
to a large relaxation time.  This expectation has been recently
substantiated by introducing a `dynamical' correlation length in terms
of a (dynamical) four points correlation 
function~\cite{chi4_1,chi4_2,chi4_3,chi4_4,chi4_5}
(see \cite{BoBiMCT,expe} for recent developments).
In this paper we confirm this line of thought through
a rigorous argument. We define a time scale 
$\tau$, and length scale $\ell$ appropriate for glassy systems and prove
that they satisfy the inequalities
\begin{eqnarray}
C_1\, \ell\le \tau\le \exp\left\{C_2\ell^d\right\}\, ,\label{eq:First}
\end{eqnarray}
where $C_{1,2}$ are two constants to be specified below, and $d$ is the 
system spatial dimension. This implies, in particular, that glassy systems
are characterized by a large length scale, increasing as the temperature
is lowered. The solution to the conundrum is that $\ell$ is 
defined in terms of  {\em point to set}
instead of {\em point to point} correlations
(which are probed, for instance, in scattering experiments).
It is worth mentioning that the definition of $\ell$ is purely statical
(it does not depend on the dynamics, as long as this satisfies a few 
conditions), and that it is closely related to the {\em Gedanken}
experiment discussed in~\cite{BiBo}. 

The physical intuition behind (\ref{eq:First}) is elementary.
The lower bound follows from the observation that in order for the system to
relax, information must be spread through at least one correlation length,
and this cannot happen quicker than ballistically. For the upper bound, one
argues that, without much harm, the system can be cut into boxes of
size $\ell$, and that, within each box, the relaxation time cannot be larger 
than exponential in the volume.

Both these arguments apply to a much larger family of models than
the ones defined on $d$-dimensional lattices. Here we consider 
systems on general factor graphs with bounded degree (see Section
\ref{sec:GraphDefinition} for a definition). 
In this case, the $\ell^d$ factor in the upper bound of Eq.~(\ref{eq:First}) 
must be replaced by the volume of a ball of radius
$\ell$ in the graph. There are two motivations
for considering such generalizations. On one hand, sparse random graphs
allow to define mean field theory, while retaining some 
features (a bounded number of neighbors and a locality structure)
of finite dimensional models. On the other hand, they appear
in a  variety of random combinatorial and constraint satisfaction problems,
ranging from coloring \cite{Coloring} to $K$-satisfiability \cite{SAT}.
In this context, one recurrent question is whether  configurations 
(or solutions) can be sampled efficiently using the Markov Chain Monte 
Carlo method. Understanding the relaxation time scale for Glauber dynamics is
relevant for this question.
In this context, point to set correlations and their relation to relaxation 
times were first considered in \cite{BergerEtAl} and subsequently
studied in \cite{MartinelliSinclair}.
In the following pages, we shall analyze one prototypical example
of such system: the $p$-spin model (also known as XOR-SAT) on random 
regular graphs.

The relation between correlations in space and time (or spatial
and temporal mixing) is indeed a well established subject in probability
and mathematical physics, with many beautiful results
(see for instance \cite{Guionnet,Martinelli} and references therein). 
There are several reasons for presenting
a new variation on this classical theme:
$(i)$ Most of the mathematical literature deals with translation invariant 
systems on finite dimensional lattices.
$(ii)$ The focus there is on `global'
characterizations of the space and time correlations, such as the
spectral gap of the Markov generator, or the log-Sobolev constant. These
are hardly accessible to experiments or numerical simulations,
and in fact are not considered in the physics literature. 
$(iii)$ The dichotomy between vanishing and not-vanishing gap (or
log-Sobolev constant) which is crucial there, is not that
important for glasses. In many such systems, the gap vanishes well above the 
glass transition due to Griffiths singularities, or to the existence of 
metastable states that are not relevant when the dynamics is
initiated with a random initial condition.
$(iv)$ The spatial mixing hypothesis used in mathematics requires
decay of correlations uniform over the boundary conditions.
This is probably too restrictive, especially in models defined on sparse 
graphs, for which the boundary of a domain can scale like its volume.

The treatment presented here is essentially self contained, and based on 
elementary combinatorial arguments (in part inspired by \cite{CombMixing}).
In Section \ref{sec:Definition} we provide our definitions of length and time
scales and state precisely our main results.
In the same Section we discuss qualitatively possible generalizations.
As an illustration of the main results we consider 
in Section \ref{sec:Examples} the $p$-spin glass model on random graphs.
In order to carry on our analysis, we prove the first rigorous
bounds on the relaxation time of a model in this class. In particular, these
imply the occurrence of a purely dynamical phase transition as the 
temperature is lowered.
Section \ref{sec:AlternativeDefinitions} contains a few alternative 
definitions of length scale and a discussion of their equivalence.
The proofs of the main results are contained in 
Sections \ref{sec:ProofLower} and
\ref{sec:ProofUpper}, with the most technical parts relegated in a series
of Appendices.
%
%
\section{Definitions and main result}
\label{sec:Definition}

\subsection{General graphical models: Equilibrium distribution}
\label{sec:GraphDefinition}

Factor graphs \cite{Factor} are a convenient language to describe
a large class of statistical mechanics models. 
A factor graph $G \equiv (V,F,E)$ is a bipartite graph with
two types of vertices: variable nodes (also called sites in the following)
$V\ni i,j,k,\dots$ and function nodes 
$F\ni a,b,c,\dots$. Edges are ordered pairs $(i,a)$, with $i\in V$ and 
$a\in F$. 
The number of variable nodes in $G$ will be denoted by $N \equiv |V|$,
and we shall identify $V=\{1,\dots,N\}$.
Given $i\in V$  (respectively $a\in F$), 
its {\em neighborhood} $\pd i$ ($\pd a$) is defined as
the set of function nodes $a$ (variable nodes $i$) 
such that $(i,a)\in E$. We assume that the graph has {\em bounded
degree}, i.e. that $|\pd i|$, $|\pd a|\le \Delta$ for some $\Delta>0$.

The {\em distance} $d_{ij}$ between two nodes $i$, $j\in V$
is the length (number of function nodes encountered along the path)
of the shortest path joining $i$ to $j$.
Given a non-negative integer $r$ and a node $i\in V$, the
{\em ball} of radius $r$ around $i$,  $\Ball_i(r)$, is the subset
of variable nodes $j$ with $d_{ij}\le r$. With a slight abuse of 
notation, $\Ball_i(r)$ will sometimes be the sub-graph induced by these
nodes. Finally, given $U\subseteq V$, we let $\pd U\subseteq V\backslash U$ 
be the subset of variable nodes at unit distance from $U$.

We deal with models with discrete variables, taking values in the
finite set $\cX$. A {\em configuration}
is a vector $x = (x_1,\dots,x_N)$, $x_i\in\cX$.
For $A\subseteq V$, we write $x_A \equiv \{x_i\, :\, i\in A\}$. 
In order to lighten the notation we shall write
$x_{\sim i,r}$ instead of $x_{V\backslash\Ball_i(r)}$ for the configuration
{\em outside} the ball of radius $r$ around $i$.

A probability distribution over configurations is defined by introducing one 
{\em compatibility function} $\psi_{\rm node}$ for each node  in $G$.
These are non-negative functions, with $\psi_a: \cX^{\pd a}\to \reals$ 
at function nodes and $\psi_i: \cX\to \reals$ at variable nodes.
We define  
\begin{eqnarray}
\mu(x) = \frac{1}{Z}\prod_{a\in F}\psi_a(x_{\pd a})
\prod_{i\in V}\psi_i(x_i)\, .\label{eq:GraphDistr}
\end{eqnarray}
Compatibility functions can also be used to define conditional 
probabilities. Let $U\subseteq V$ and $F(U)\subseteq F$ be the subset 
of function nodes having at least one neighbor in $U$.
If $y =(y_1,\dots,y_N)$ is a configuration, we define
\begin{eqnarray}
\mu_U^y(x) = 
\frac{1}{Z_U^y}\prod_{a\in F(U)}\psi_a(x_{\pd a})
\prod_{i\in U}\psi_i(x_i)\, ,\label{eq:ConditionalDef}
\end{eqnarray}
if $x$ coincides with $y$ on $U^{\rm c}\equiv V\backslash U$ 
(i.e. $x_{U^{\rm c}}= y_{U^{\rm c}}$) and $\mu_U^y(x) = 0$
otherwise. The normalization constant $Z_U^y$ ensures that
$\sum_x\mu_U^y(x)=1$.

Since we allow for vanishing compatibility functions (hard core 
interactions), the above expressions could be {\em a priori} ill defined.
In order to avoid this, and to simplify our treatment, we shall restrict
ourselves to {\em permissive} interactions
(the definition we give here is very close to the one of \cite{CombMixing}). 
This means that for  each site $i$ there exists $x^*_i\in\cX$ 
such that, given any non-empty $U\subseteq V$, if  $x_i=x^*_i$ 
for each $i\in U$, then the 
right hand side of Eq.~(\ref{eq:ConditionalDef}) is strictly positive 
regardless of the assignment of $x$ on $\pd U$. 
In particular, this must be the case
when the subset is a single vertex $U=\{i\}$. We denote by 
$\mu_0>0$ a lower bound on the conditional probability for the
corresponding state to be $x_i^*$. In other words, we ask that
$\mu_i^{y}(x_i^*) \ge \mu_0$ for all $i$ and $y$.
%
%
\subsection{Glauber dynamics}
\label{sec:DynamicsDefinition}

The dynamics is specified as a single spin flip, continuous--time
Markov process, irreducible, aperiodic and satisfying detailed 
balance with respect to the distribution (\ref{eq:GraphDistr}).
More precisely, for each variable node $i$, a set of transition rates 
$\kappa_i^x(\xi)\ge 0$, with $\xi\in\cX$ and 
$\sum_\xi\kappa_i^x(\xi) = 1$ is specified.
Each variable node $i\in V$ is associated to a clock, whose ringing times
are distributed according to independent rate--one Poisson processes.
When the clock at site $i$ rings, a new value $\xi\in\cX$ is drawn from the 
distribution $\kappa_i^x(\xi)$, $x$ being the current configuration. The new
configuration $x'$ coincides everywhere with $x$ but on $i$,
where $x'_i=\xi$. In order to verify detailed balance, the following condition
must be satisfied by the transition rates:
\begin{eqnarray}
\mu(x)\, \kappa_i^x(x'_i) = \mu(x') \, \kappa_i^{x'}(x_i)\, .
\end{eqnarray}
for any two configurations $x$ and $x'$ that coincide everywhere but on $i$.

In our treatment we shall make two assumptions on the transition rates.
\begin{itemize}
\item {\em Locality}. The transition rates $\kappa_i^x(\,\cdot\,)$
depend on the current configuration $x$ only through 
$x_j$, with $d_{ij}\le 1$. Although its precise form could be 
modified (and somewhat relaxed), this is a crucial physical 
requirement: the Markov dynamics must be local with respect to the underlying
graph $G$.
\item {\em Permissivity}. Let $x^*_i$ the state permitted at node $i$
regardless of the configuration at $V\backslash \{i\}$. Then, there exists
$\kappa_0>0$ such that $\kappa^x_i(x^*_i)\ge \kappa_0$ independently of 
$i$ and $x$. This assumption can somewhat be relaxed at the expense
of some technical difficulties.
\end{itemize}

A well known example of transition rates satisfying these conditions
is given by the so-called `heat-bath rule':
\begin{eqnarray}
\kappa^x_i(\xi) = \mu^x_i(\xi)\, .
\end{eqnarray}

It will be also useful to define the Markov dynamics on a 
subset of the vertices $U\subseteq V$, with boundary condition 
$y\in\cX^V$. By this we mean that the initial configuration agrees
with $y$ on $V\backslash U$, and that variables outside $U$ are `frozen':
each time the clock rings at a site $i\in V\backslash U$, the
configuration is left unchanged.

We shall sometimes use the notation $\<\, \cdot\, \>$ for averages 
with respect to the equilibrium distribution $\mu$ or the Markov process
defined above, with initial condition distributed according to $\mu$. We also
denote by $\<\,\cdot\,\>_{U}^y$ averages with respect to the
subset in $U$ with boundary $y$ (i.e. either with respect to the 
distribution $\mu^y_U$ or with respect to the Markov process on $U$
with initial condition distributed according to $\mu^y_U$).
%
%
\subsection{Length and time scales and their relation}
\label{sec:LengthTime}

Both our definitions of length and time scales depend on 
a parameter $\ve$. This should be thought of as some fixed small
number (let's say $0.01$), that provides a cut-off for
distinguishing `highly correlated' from `weakly correlated'
degrees of freedom. The physical idea is that, near a glass transition,
the order of magnitude of the resulting length and time scales should be 
roughly independent of $\ve$ as long as this is smaller than a 
characteristic value (the `Edwards-Anderson' or `ergodicity breaking' 
parameter).

Consider a vertex $i$ in $G$. Let $f(x_i)$ be a function of the variable at
$i$, and $F(x_{\sim i,r})$ a function of the variables whose distance from $i$
are larger than $r$. The covariance 
$\<f(x_i)F(x_{\sim i,r})\>-\<f(x_i)\>\<F(x_{\sim i,r})\>$ measures the 
degree of correlation between the observables $f$ and $F$. 
In order to quantify the degree of correlation of $x_i$ and $x_{\sim i,r}$,
it makes sense to consider the `most correlated' observables and define
\begin{eqnarray}
G_i(r) \equiv \sup_{f,F}\,
\Big|\<f(x_i)F(x_{\sim i,r})\>-\<f(x_i)\>\<F(x_{\sim i,r})\>\Big|\, ,
\label{eq:SpatialCorrDef}
\end{eqnarray}
where the $\sup$ is taken over all functions such that 
$|f(x_i)|, |F(x_{\sim i,r})|\le 1$ for any $x$.
The correlation length $\ell_i(\ve)$ of vertex $i$ is defined as the smallest
integer $\ell$ such that $G_i(r)\le \ve$ for all $r\ge\ell$.
In formulae
\begin{eqnarray}
\ell_i(\ve) \equiv \min\{\ell\ge 0\;\;
\mbox{ s.t. }\;\;G_i(r)\le \ve\;\; \forall 
\;r\ge\ell\}\,.\label{eq:ellDef}
\end{eqnarray}
If no such $\ell$ exists, $\ell_i(\ve)$ is set by convention to the 
maximum distance from $i$ of a vertex in $G$.

The time scale definition is completely analogous to the above one.
We let 
\begin{eqnarray}
C_i(t) =\sup_{f}\left|
  \<f(x_i(0))f(x_i(t))\>-\<f(x_i(0))\>\<f(x_i(t))\>\right|\, ,
\end{eqnarray}
the $\sup$ being taken over functions of the variable at $i$
such that $|f(x_i)|\le 1$ for any $x_i$. Then we let
\begin{eqnarray}
\tau_i(\ve) = \inf \{\tau\ge 0\;\;\mbox{ s.t. }\;\;C_i(t)\le \ve\;\; \forall 
\;t\ge\tau\}\,.
\end{eqnarray}
This expression is well defined, and the resulting $\tau_i(\ve)$ is
always finite. In fact, if we define
$C_i^f(t)\equiv \<f(x_i(0))f(x_i(t))\>-\<f(x_i(0))\>\<f(x_i(t))\>$,
 the spectral representation of the transition 
probabilities~\cite{AldousFill}, 
implies $C_i^f(t)= \sum_{l=2}^n e^{- \lambda_l t} B_l(i,f)$.
Here $0 < \lambda_2 \le \lambda_3 \le \dots$ are the eigenvalues of the 
Markov generator, the $B_l$ are non-negative coefficients, and 
$n$ is the number of configurations of the system. 
In consequence this correlation function is positive, decreasing and has
a vanishing limit when $t \to \infty$.

Both definitions given here admit several essentially equivalent 
variants which can be helpful depending on the circumstances;
we shall discuss some of them in Sec.~\ref{sec:AlternativeDefinitions}.
We can now state our main result, whose proof can be found in 
Secs.~\ref{sec:ProofLower} and \ref{sec:ProofUpper}.

\begin{thm}\label{thm:Main}
Under the hypothesis presented in this Section
\begin{eqnarray}
C_1\ell_i(|\cX| \sqrt{2 \ve})
\le\tau_i(\ve)\le 1 +  
A\,\exp\left\{C_2 |\Ball_i(\ell_i(\ve/2))|\right\}
\ ,
\end{eqnarray}
where  $A=\log\left(\frac{4}{\ve}\right)$, $C_1 = 1/2e\Delta^2$,
$C_2 = -\log(\kappa_0 (1-e^{-1}))$, 
and the lower bound holds under the
assumption $\ell_i(|\cX| \sqrt{2 \ve}) > \log_2 (2/\ve) $.
\end{thm}
Let us stress that Ref.~\cite{BergerEtAl} (Theorem 1.5) 
proves a very general result that is closely related to the above lower bound. 
Despite the fact that both the statements and the proofs are quite similar, 
we think that our formulation can be better suited for physics applications. 
The authors of \cite{BergerEtAl} provide in fact a bound
on the spectral gap, a quantity that is hardly accessible in experiments
or simulations. Further it does not correspond to an interesting time scale 
when studying the glass transition. In fact, the relaxation time defined as
the inverse spectral gap becomes exponential in the system size well above 
the dynamical glass transition (roughly speaking, at the appearence of the 
first metastable states).
This can be shown, for instance, in the example of Section~\ref{sec:Examples}.
%
%
\subsection{Problems and generalizations}

Let us briefly discuss Theorem \ref{thm:Main} and a few related
research directions. As shown in Section
\ref{sec:Examples}, this theorem can be only marginally improved in 
the general setting
described above. It would therefore be interesting to determine
additional conditions under which the upper/lower bounds are in fact closer
to the actual correlation time. 
 
Recently there has been a considerable interest in 
a `dynamical' length scale $\xi_4$ defined through four point correlation 
functions \cite{expe}. It would be interesting to understand 
whether a general relation holds between $\xi_4$, and the length 
$\ell$ defined in Eq.~(\ref{eq:ellDef}). Within mean field models
undergoing a discontinuous glass transition,  it has been argued that they 
are in fact closely related \cite{NostroLettera,notre_long}. 

The notion of `growing length scale' is useful also when the
initial condition for the Markov dynamics is not drawn from the 
equilibrium measure\footnote{The question was posed to us by Giulio Biroli.} 
$\mu(x)$. Of particular interest is the case 
of a uniformly random initial condition $x\in\cX^N$
(a `quench from infinite temperature'). The 
definition (\ref{eq:ellDef}) can be generalized to this setting
if  the expectations in Eq.~(\ref{eq:SpatialCorrDef}) are taken with respect 
to the measure at some time $t\ge 0$. The resulting length will depend on time:
$\ell_i(\ve,t)$, and is expected to increase with $t$, starting from 
$\ell_i(\ve,0) =0$ until it reaches the equilibrium value
$\ell_i(\ve,\infty) = \ell_i(\ve)$. The disagreement percolation technique
used to prove the lower bound in Theorem \ref{thm:Main}
can be generalized to this case, cf. Section~\ref{sec:ProofLower}, to show
that $\ell_i(\ve,t)\le \tilde{C}\, t$.

Finally, it would be interesting to consider more realistic 
models for glasses, e.g. off-lattice particle systems with 
Langevin dynamics. We guess that similar ideas to the ones 
proposed here can be useful in that case, although at the expenses of
several technical difficulties.
%
%
\section{Alternative definitions of length scale}
\label{sec:AlternativeDefinitions}

The basic physical idea in the definition of length scale is to
look at correlations between a point (a vertex) and the whole set
of variables at distance larger than $r$ from it.
Equation (\ref{eq:SpatialCorrDef}) provides one measure of these correlations.
Here we shall define four alternative measures $G^{(n)}_i(r)$,
$n=1,\dots,4$. For any such measure, we can define a length scale
exactly as in Section \ref{sec:LengthTime},
\begin{eqnarray}
\ell^{(n)}_i(\ve) \equiv \min\{\ell\ge 0\;\;\mbox{ s.t. }\;\;G^{(n)}_i(r)\le 
\ve\;\; \forall  \;r\ge\ell\}\,.\label{eq:AlternativeLengths}
\end{eqnarray}
The question which we shall address shortly is to what extent these
definitions are equivalent.

The first two definitions express the idea that two random variables
are weakly correlated if their joint distribution is (approximately) 
factorized. With an abuse of notation we denote by $\mu(x_i,x_{\sim i,r})$
the joint distribution of $x_i$ and $x_{\sim i,r}$, 
and by $\mu(x_i)$, $\mu(x_{\sim i,r})$ their marginal distributions.
Then we define
\begin{eqnarray}
G^{(1)}_i (r) &\equiv & \sup_{x_i,x_i'}\sum_{x_{\sim i,r}}
\Big|\mu(x_{\sim i,r}|x_i)-\mu(x_{\sim i,r}|x_i')\Big|\, ,\\
G^{(2)}_i(r) &\equiv & \sum_{x_i,x_{\sim i,r}}\Big|
\mu(x_i,x_{\sim i,r})-\mu(x_i)\mu(x_{\sim i,r})\Big|\, .
\end{eqnarray} 

One inconvenient of the definition of $G_i(r)$, 
cf. Eq.~(\ref{eq:SpatialCorrDef}), as well as of $G^{(1)}_i(r)$ and 
$G^{(2)}_i(r)$ is that they are difficult  to evaluate. 
Equation (\ref{eq:SpatialCorrDef}), for instance, requires the optimization
with respect to $F$ which is, in general, a function of $\Theta(N)$ variables.
Given a function $f$ of the variable at $i$, we let $\fp$ be the function 
of $x_{\sim i,r}$ obtained by taking the expectation of $f$ with respect to
the conditional distribution corresponding to the boundary condition
 $x_{\sim i,r}$ outside $\Ball_i(r)$. In formulae:
$\fp(x_{\sim i,r})= \<f(x_i)\>^x_{\Ball_i(r)}$. We then define
\begin{eqnarray}
G^{(3)}_i(r) = \sup_f \Big|\<f(x_i)\fp(x_{\sim i,r})\>-
\<f(x_i)\> \<\fp(x_{\sim i,r})\>\Big|\, ,
\end{eqnarray}
where, again, the $\sup$ is taken over the functions $f$ such that
$|f(x_i)|\le 1$ for any $x$. 
 Note that one can always decompose
an arbitrary function $f(x_i)$ in terms of the $|\cX|$ indicator functions
$f_\xi(x_i) = \ind(x_i=\xi)$, hence $G^{(3)}_i(r)$ can be computed with
a finite number of covariance estimations. The correlation
function $G^{(3)}_i(r)$ was already used in \cite{MartinelliSinclair}
to bound the spectral gap of Ising and hard core models on trees.

A suggestive interpretation of the last definition is provided by the
following procedure. Generate a reference configuration $x$ according
to the distribution $\mu$, and evaluate $f(x_i)$ on it. Then `freeze' 
everything is outside $\Ball_i(r)$ and generate a new configuration
$x'_i$ inside, according to the conditional distribution
$\mu_{\Ball_i(r)}^x$ (i.e. with the boundary condition given by the frozen
variables). Evaluate $f(x'_i)$ on the new configuration.
The correlation function $G^{(3)}_i(r)$ is (the $\sup$ over $f$ of)
the covariance between $f(x_i)$ and $f(x'_i)$. This procedure was indeed
discussed in~\cite{BiBo}.

A last definition consists in considering the mutual information 
\cite{CoverThomas} between $x_i$ and $x_{\sim i,r}$:
\begin{eqnarray}
G^{(4)}_i(r) \equiv I(X_i;X_{\sim i,r})\, .
\end{eqnarray}
Recall that, given two discrete random variables $X$ and $Y$ with distribution 
$p(x,y)$, their mutual information is defined as 
$I(X;Y) = \sum_{x,y} p(x,y)\log\frac{p(x,y)}{p(x)p(y)}$.  A pleasing
interpretation follows from the general principles of information theory.
Suppose that a configuration $x$ is generated according to the distribution
$\mu$, but only $x_{\sim i,r}$ is revealed to you. In a very precise sense,
$G_i^{(4)}(r)$ measures how much information you would have about $x_i$.

It turns out that the length scales extracted from $G^{(1)}(r)$,
\dots, $G^{(4)}_i(r)$ convey essentially the same information as 
$\ell_i(\ve)$. More precisely, they differ only by a redefinition of 
the parameter $\ve$ and a rescaling.
\begin{propo}\label{propo:CorrEquiv}
Let $\ell^{(n)}_i(\ve)$, $n=1,\dots,4$  be defined as in
Eq.~(\ref{eq:AlternativeLengths}), and $\ps \equiv \min_{x_i\in\cX}\mu(x_i)$. 
Then
\begin{align}
\ell^{(2)}_i(|\cX|\ve)&\le\ell_i(\ve)\;\, \le \ell^{(2)}_i(\ve)\, ,\\
\ell^{(2)}_i(\ve)&\le\ell^{(1)}_i(\ve)\le \ell^{(2)}_i(\ps\ve/2)\, ,\\
\ell^{(2)}_i(|\cX|\sqrt{\ve})&\le
\ell^{(3)}_i(\ve)\le \ell_i(\ve)\, ,\\
\ell^{(2)}_i(\sqrt{2\ve})&\le
\ell^{(4)}_i(\ve)\le \ell^{(2)}_i\left(\frac{\ve\ps}{1-\ps}\right)\, .
\end{align}
\end{propo}
Under the permissivity assumptions, we have $\mu_*\ge \mu_0>0$
and the changes of argument in the above expressions amount therefore
to a finite rescaling in $\ve$.
The proof of this proposition can be found in App.~\ref{app:CorrEquiv}.
%
%
\section{A mean field example: the $p$-spin model on random (hyper)graphs}
\label{sec:Examples}

In this Section we discuss a particular example exhibiting
glassy behavior. Our objective is twofold. First of all, we want to 
check to what extent Theorem \ref{thm:Main} can be improved over. Second, 
we want to show that the idea of diverging correlation can be of
relevance even for mean field systems, as soon as the underlying factor
graph is sparse. 

We consider a model of $N$ Ising spins 
$\sigma = (\sigma_1,\dots,\sigma_N)$, $\sigma_i\in\{+1,-1\}$
(for historical reasons, we use here $\sigma_i$ instead of $x_i$ to
denote the $i$-th variable), 
defined through the Boltzmann distribution
\begin{eqnarray}
\mu(\sigma) & = &\frac{1}{Z(\beta)}\, e^{-\beta E(\sigma)}\, ,
\label{eq:PspinBoltzmann}\\
E(\sigma) & = & -\sum_{a=1}^{M}J_a\prod_{i\in\partial a}\sigma_i
\equiv -\sum_{a=1}^{M}J_a\sigma_{i_1(a)}\cdots \sigma_{i_p(a)}\, .
\label{eq:PspinEnergy}
\end{eqnarray}
Here we think of $i\in\{1,\dots,N\}$ as the variable nodes of a factor
graph $G$, and of $a\in\{1,\dots,M\}$ as its function nodes.
Further, we assume $G$ to be a random factor 
graph with degree
$l$ at variable nodes, and $p$ at function 
nodes\footnote{More precisely $G$ is distributed according
to the corresponding {\em configuration model}. To sample a factor graph
from this ensemble, each of the $N$ variable nodes is attributed $l$
sockets, and each of the $M$ function nodes, $p$ sockets. 
The $Nl=Mp$ sockets on the two sides are then matched according to a 
uniformly random permutation over $Nl$ elements.
Multiple edges are removed, if they occur an even number of times, 
and replaced by a single edge, in the opposite case.}. A random 
factor graph from this ensemble will also be referred to as
a `random $l$-regular hypergraph' (function nodes being identified with 
hyperedges joining the neighboring variable nodes).
 Finally,
$\beta = 1/T$ is the inverse temperature and the $J_{a}$'s are i.i.d.
random variables, uniform in $\{+1,-1\}$. 

In the following it will be always understood that 
$p,l\ge 3$. It is in fact expected that for $p=2$ the model undergoes
a spin glass transition without a dynamical phase transition, and for 
$l=2$, no phase transition at all occurs at finite temperature.

As the variables $\sigma_i$ can only take two opposite values, the
general definitions given in Sec.~\ref{sec:LengthTime} simplify somewhat.
For instance
\begin{eqnarray}
C_i(t) &=& \<\sigma_i(0) \sigma_i(t) \> - \<\sigma_i\>^2\, , \\
G_i^{(3)}(r) &=& \left|\< \sigma_i \, \<\sigma_i \>_{\Ball_i(r)}^\sigma \> 
- \<\sigma_i\>^2 \right| \, .
\end{eqnarray}

The phase diagram of this model is expected to be characterized by two phase
transitions~\cite{pspin}:
a dynamical phase transition at temperature $T_{\rm d}$,
and a statical one at $T_{\rm c}<T_{\rm d}$
(see also \cite{NostroLettera,notre_long}
for a detailed study of the dynamics in the case of uniformly drawn random 
hypergraphs). At high temperature 
$T>T_{\rm d}$, Glauber dynamics is fast and the relaxation time $\tau$ 
does not depend on the system size $N$.
Note that, in the thermodynamic limit, any finite 
neighborhood of any spin is a regular hypertree with high probability.
Therefore as far as the $\tau_i$'s are finite, they converge in probability 
to a  deterministic value $\tau$.

Below $T_{\rm d}$, the $\tau_i$'s become instead exponentially large in $N$. 
Heuristic estimates on the exponential rate can be expressed in terms of 
free-energy barriers, computed from the so-called quenched 
potential~\cite{potential,Silvio_Kac}. In \cite{Cristina},
it was shown that the spectral gap is exponentially 
small\footnote{One interesting feature is that the temperature $T_{\rm gap}$
below which the gap becomes exponentially small, is higher than $T_{\rm d}$.
In the temperature range $T_{\rm d}<T<T_{\rm gap}$, the slow modes 
correspond to metastable states that are `not seen' if the dynamics starts from
a random or equilibrated initial condition.} at 
low enough temperature for some values of $p$ and $l$.

Remarkably, the 
free energy remains analytic in the interval  $(T_{\rm c},\infty)$, and 
a true thermodynamic phase transition takes place only at $T_{\rm c}$.
If $p \ge l$, $T_{\rm c} = 0$, and there is no thermodynamic phase transition.

Assuming that this picture is correct (below we shall partially confirm it),  
Theorem \ref{thm:Main} implies that the correlation lengths 
$\ell_i$ are finite for $T>T_{\rm d}$.
On the other hand, for $T<T_{\rm d}$, the $\ell_i$'s are necessarily divergent 
in the system size. Since the size of a ball of radius $r$
is bounded as $|\Ball_i(r)|\le l(p-1)^r(l-1)^r$, a moment of thought shows
that $\ell\ge C \log N$. On the other hand, $\ell$ cannot be larger than the
graph diameter, whence $\ell = \Theta(\log N)$.

In the high temperature phase $T>T_{\rm d}$, the correlation function 
$G^{(3)}(r)$ can be computed recursively, using a non-rigorous approach
that exploits the locally tree-like structure of the factor graph.
Let us sketch the procedure here and refer to \cite{notre_long,reconstruction} 
for a detailed description.
We consider a rooted tree factor graph with $R$ generations $\Tree_*(R)$
defined as follows. For $R=0$, $\Tree_*(R)$ is a single node (the root). 
For any $R\ge 0$,
one first defines $\Tree_*(R)'$ by joining $(l-1)$ copies of $\Tree_*(R)$
at the root, and then joins $(p-1)$ copies of $\Tree_*(R)'$ to a common 
function node $a$ of degree $p$ to obtain $\Tree_*(R+1)$. The remaining 
variable node adjacent to $a$ is the new root. One then generate 
a configuration $\sigma$ on this graph according to the Boltzmann weight 
(\ref{eq:PspinBoltzmann}), `freezes' it from generation $r$ on,
and consider the conditional $\mu_{i,r}(\,\cdot\,)
=\mu(\sigma_i=\,\cdot\,|\sigma_{\sim i,r})$,  $i$ being the root node.
When $\sigma_{\sim i,r}$ is generated randomly according to the
above procedure,  $\mu_{i,r}(\, \cdot\, )$ can be considered as a random 
variable. Knowing its distribution allows to compute $G^{(3)}(r)$
(it turns out that this distribution does not depend upon $R$).
Thanks to the tree structure, a recursive distributional equation can be 
written for $\mu_{i,r}(\, \cdot\, )$, and solved
numerically using a sampling (`population dynamics') technique.

\begin{figure}
\includegraphics[width=8.5cm]{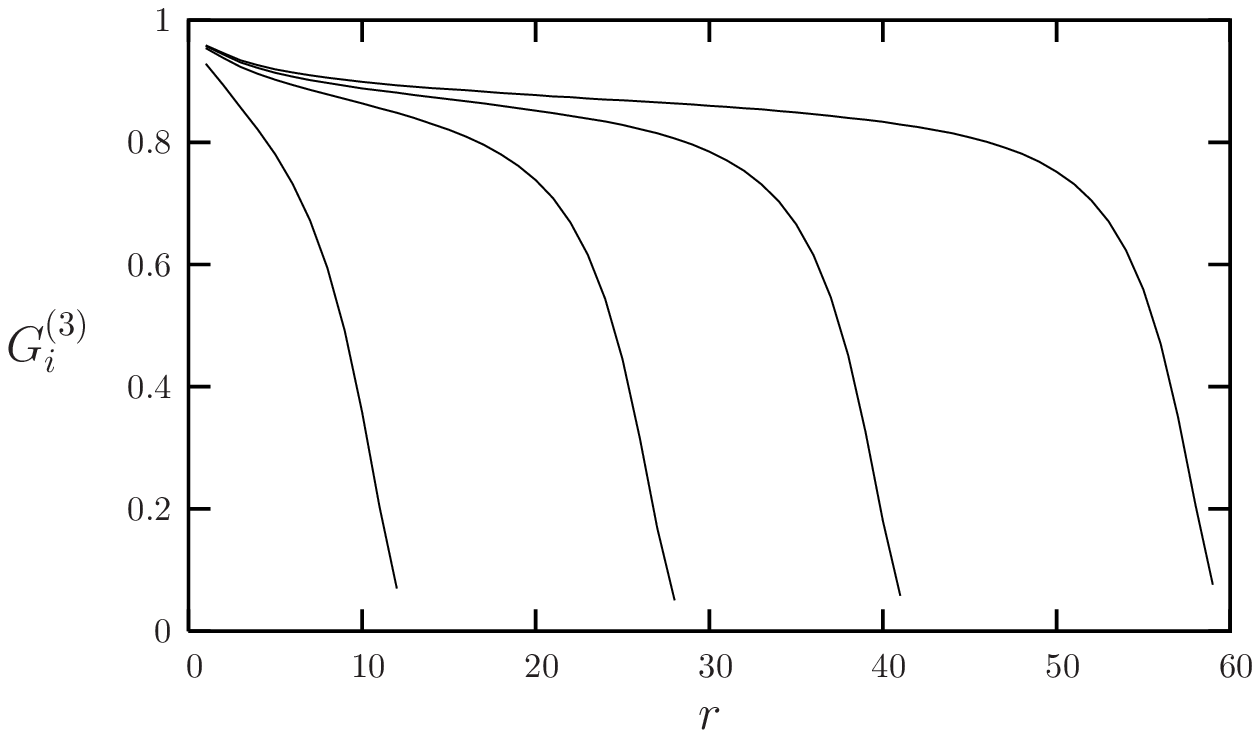}
\includegraphics[width=8.5cm]{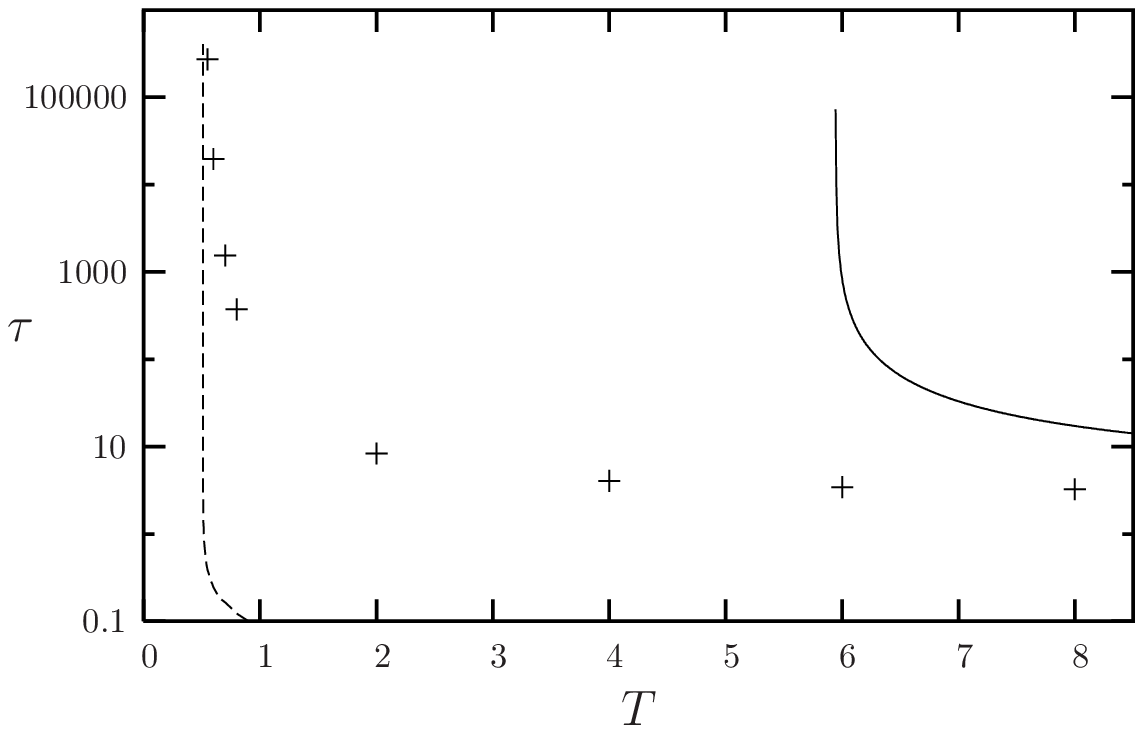}
\caption{Numerical results for the $3$-spin model on random $3$-regular 
hypergraphs
at high temperature. Left: the spatial correlation function $G_i^{(3)}(r)$, 
from left to right $T=0.6$, $0.53$, $0.52$, $0.515$. Right: correlation times. 
Symbols are obtained by Monte Carlo simulations of the heat bath dynamics, 
the solid line is the rigorous upper bound of Proposition 
\ref{propo:TimePspin}, the dashed line corresponds to the
lower bound from Theorem \ref{thm:Main}.}
\label{fig:highT}
\end{figure}

In Fig.~\ref{fig:highT}, left frame, we plot the results of
such a computation for $p=l=3$ and a few values of $T$. 
As the temperature is lowered towards $T_{\rm d}$, $G^{(3)}(r)$ develops 
a plateau of length diverging as $(T-T_{\rm d})^{-1/2}$. 
As a consequence, for any $\ve$ smaller than the value of 
$G^{(3)}(r)$ on the plateau, $\ell^{(3)}(\ve)\sim (T-T_{\rm d})^{-1/2}$.
In order to check whether the lower bound of Theorem \ref{thm:Main}
is optimal, we estimated the time $\tau$ from 
Monte Carlo simulations of the heat bath dynamics on large ($N=10^6$) samples.
We averaged over several samples and
checked that $\tau_i$ is approximately independent of $i$. 
Our data are presented in the right frame of Fig.~\ref{fig:highT}
and compared with the lower bound of Theorem \ref{thm:Main}
with $\ell^{(3)}$ evaluated via the recursive method.
The correlation time turns out to diverge algebraically 
at $T_{\rm d}$: $\tau\sim (T-T_{\rm d})^{-\gamma}$.
Fitting the data, we get $\gamma \approx 3.2 >1/2$.

\begin{figure}
\center{\includegraphics[width=8.5cm]{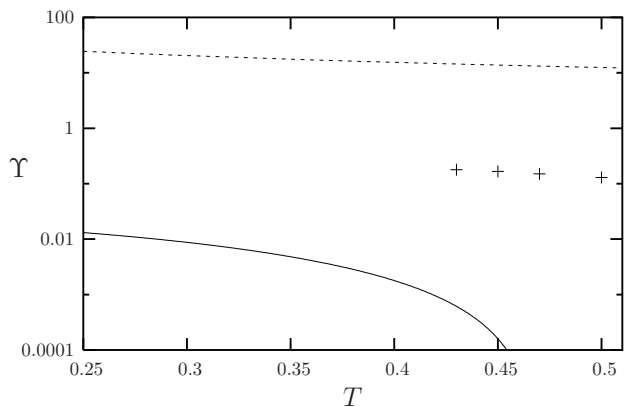}}
\caption{Correlation times for the $3$-spin model on random $3$-regular 
graphs at low temperature.
Here we plot $\Upsilon = (\log \overline\tau)/N$ versus temperature
($\overline\tau$ being uniform average of $\tau_i$ over the site $i$).
The continuous line is the rigorous lower bound from Proposition 
\ref{propo:TimePspin}. The dashed line is the upper bound 
from Theorem \ref{thm:Main}.
Symbols correspond to numerical results from Monte Carlo simulations.}
\label{fig:lowT}
\end{figure}

In Fig.~\ref{fig:lowT} we plot the exponential growth rate of the
correlation time obtained by extensive numerical simulations on small
($N=100$) systems (see also~\cite{barrieres_Enzo} for an analogous study in the
fully connected limit). 
As above, we consider here the case $l=3$, $p=3$.
This is compared with a rigorous lower bound stated in 
Proposition \ref{propo:TimePspin} below, and with the upper bound
obtained from Theorem \ref{thm:Main}, using $|\Ball_i(r)|\le N$.

The above discussion of the behavior of correlation times was largely based 
on the analogy with the fully connected spherical model
\cite{DynamicsReview}, and an on heuristic arguments.
Here we confirm rigorously several elements of this picture. For the
sake of definiteness, we shall consider Glauber dynamics with the heat
bath rule
\begin{eqnarray}
\kappa_i^{\sigma}(\sigma_i) = \frac{1}{2}\,
\left(1+\sigma_i\tanh\beta h_i(\sigma)\right)\, ,
\end{eqnarray}
where $h_i(\sigma) = 
\sum_{a \in\partial i}J_a\prod_{j\in\partial a\backslash i}\sigma_j$ 
is the `local field' acting on the $i$-th spin. 
Then we have the following result, whose proof is deferred
to Appendices \ref{sec:HighTPspin} and \ref{sec:LowTPspin}.
\begin{propo}\label{propo:TimePspin}
Let $\tau_i(\ve)$ be the correlation time for  spin $\sigma_i$ in a $p$-spin 
model on a random $l$-regular hypergraph with $p$, $l\ge 3$. Let 
$T^{\rm fast}_{p,l} = ({\rm arctanh}(1/l(p-1)))^{-1}$, and 
$T^{\rm barr}_{p,l}$, $T^{\rm ann}_{p,l}$ 
be the temperatures defined in Appendix~\ref{sec:LowTPspin}. 

If $T>T^{\rm fast}_{p,l}$, then
$\tau_i(\ve) \le (1/\kappa) \log(1/(\kappa \ve))$, where 
$\kappa \equiv (1-l(p-1)\tanh\beta)$.

If $T^{\rm ann}_{p,l}<T<T^{\rm barr}_{p,l}$, then there
exists constants $q_*$ and $\Upsilon>0$ such that, for any $1/4>\delta>0$,
$\tau_i(\ve)\ge e^{N[\Upsilon-\delta]}$ for at least $N(q_*-\delta-\ve)$
spins $\sigma_i$ with high probability.
\end{propo}
The analysis in Appendix  \ref{sec:LowTPspin}  
also provides rather explicit expressions 
for the temperatures $T^{\rm barr}_{p,l}$,
$T^{\rm ann}_{p,l}$ as well as for $q_*$ and $\Upsilon$. 
The numerical values of some of these constants are reported in
Table~\ref{tab:temperatures} for a few values of $p$ and $l$.

\begin{table}\label{tab:temperatures}
\begin{center}
\begin{tabular}{| c | c | c | c | c | c |}
\hline
$p$ & $l$ & $T_{\rm ann}$ & $T_{\rm barr}$ & $T_{\rm d}$ & $T_{\rm fast}$ \\
\hline
\hline
3 & 3 &0 &0.470124 &0.510 &5.944027 \\
\hline
3 & 4 &0.854138 &0.687684 &0.753 &7.958158 \\
\hline
3 & 5 &1.113214 &0.849507 &0.935 &9.966577 \\
\hline
\hline
4 & 3 &0 &0.376808 &0.410 &8.962840 \\
\hline
4 & 4 &0 &0.575513 &0.625 &11.972171 \\
\hline
4 & 5 &0.771325 &0.724693 &0.785 &14.977751 \\
\hline
\hline
\end{tabular}
\end{center}
\caption{Various characteristic temperatures for the  $p$-spin model
on random $l$-regular hypergraphs: 
$T_{\rm ann}$ is an upper bound on the static
transition temperature $T_{\rm c}$; $T_{\rm d}$ is the dynamical transition
obtained by a cavity calculation; 
$T_{\rm barr}$ and $T_{\rm fast}$ are, respectively,
lower and upper bounds on $T_{\rm d}$.}
\end{table}

We are now in position to discuss to what extent Theorem \ref{thm:Main}
can be improved (here we focus on the large $\ell$, large $\tau$
behavior) without loosing in the generality of its hypotheses. 
In the high temperature phase 
$\tau\sim(T-T_{\rm d})^{-\gamma}$, while $\ell\sim(T-T_{\rm d})^{-1/2}$,
and the analogy with fully connected models suggests $\gamma\ge 1$ quite
generically~\cite{DynamicsReview}. Therefore we expect that the lower bound
can be improved at most to $\tau\ge \ell^{\zeta}$ with, probably,
$\zeta =2$.  

At low temperature $\tau = \exp\{\Theta(N)\}$ while
Theorem \ref{thm:Main} implies $\tau \le \exp\{C N\}$. This is optimal apart 
from a possible improvement in the exponential rate. 
Our conclusion is that, without further assumptions on the system,
Theorem \ref{thm:Main} can be improved at most to
$C_1\ell_i^\zeta\le\tau_i\le\exp\{C_2|\Ball_i(\ell)|\}$.

Let us finally observe that Proposition \ref{propo:TimePspin},
together with Lemma \ref{lemma:PartF} imply that the model 
(\ref{eq:PspinEnergy}) undergoes a purely dynamical phase transition. 
This is the first time such a behavior is proved for a model of this
family (Glauber dynamics on a sparse graph spin model).
In finite dimensional models, we expect that such a sharp dynamical
transition cannot occur. Nevertheless, it would be interesting to understand
whether the correlation time of glassy systems generically
undergoes a crossover from a polynomial ($\tau\sim \ell^z$) to an exponential
($\tau\sim\exp\{\Theta(\ell^\psi)\}$) relation with the correlation length. 
%
%
\section{Proof of the lower bound in Theorem \ref{thm:Main}}
\label{sec:ProofLower}

For pedagogical reasons we first recall some definitions and well-known 
facts from probability theory, 
that will be repeatedly used in this and the following
proofs. Consider a finite set $S$ and two probability measures $p^{(1)}$, 
$p^{(2)}$ defined on it. The total variation distance between these 
two measures is defined as
\begin{equation}
||p^{(1)}-p^{(2)}||_\sTV\equiv\frac{1}{2}\sum_x|p^{(1)}(x)-p^{(2)}(x)| \ . 
\end{equation}
One easily checks that this defines indeed a distance, in particular it 
vanishes if and only if $p^{(1)}=p^{(2)}$, and that the two following
characterizations are equivalent:
\begin{equation}
||p^{(1)}-p^{(2)}||_\sTV = \underset{T \subset S}{\max}|p^{(1)}(T)-p^{(2)}(T)|
= 1 - \sum_x \overline{p}(x) \ ,
\end{equation}
where we defined $\overline{p}(x)=\min[p^{(1)}(x),p^{(2)}(x)]$.

Another very useful form of the total variation distance is defined in terms
of couplings of the two measures, that is to say joint distributions
$q(x_1,x_2)$ whose marginals are $p^{(1)}$ and $p^{(2)}$:
$\sum_{x_2} q(x_1,x_2) = p^{(1)}(x_1)$ and 
$\sum_{x_1} q(x_1,x_2) = p^{(2)}(x_2)$.

Considering the random variable $(X_1,X_2) \in S^2$ drawn from such a coupling,
one can easily prove that
\begin{equation}
||p^{(1)}-p^{(2)}||_\sTV \le \prob[ X_1 \neq X_2 ] \ .
\end{equation}
Moreover one can construct an optimal (or greedy) coupling that achieves the 
bound. 
We thus have
\begin{equation}
||p^{(1)}-p^{(2)}||_\sTV = \underset{q}{\min} \ \prob[ X_1 \neq X_2 ] \ .
\end{equation}

We can now start the proof of the lower bound in Theorem~\ref{thm:Main}
(a presentation in a restricted setting is 
provided in \cite{notre_long}).
This is based on a `disagreement percolation' argument first
used in \cite{VanDenBerg} (for  recent applications, see \cite{BergerEtAl,HaSi}).
More precisely, for a given variable node $i$ and a positive integer $r$,
we construct a Markov process $(x^{(1)}(t),x^{(2)}(t))$ on 
$\cX^N \times \cX^N$ in the following way:
\begin{itemize}
\item[$\bullet$] at the initial time $t=0$, a configuration $x(0)$ is drawn
from the law $\mu$, and imposed to the two trajectories, 
$x^{(1)}(0),x^{(2)}(0)=x(0)$.
\item[$\bullet$] Each variable node owns an independent rate-one Poisson clock.
When the clock at $j$ rings, say at time $t$:
\begin{itemize}
\item if $j \notin \Ball_i(r)$, $x_j^{(2)}(t)$ is replaced by $\xi$, drawn
from $\kappa_j^{x^{(2)}(t)}(\xi)$, and $x^{(1)}$ remains unchanged.
\item if $j \in \Ball_i(r)$, we draw $(\xi_1,\xi_2)$ from the greedy coupling
between $\kappa_j^{x^{(1)}(t)}$ and $\kappa_j^{x^{(2)}(t)}$, and replace
$(x_j^{(1)}(t),x_j^{(2)}(t))$ by $(\xi_1,\xi_2)$.
\end{itemize}
\end{itemize}
$(x^{(1)}(t),x^{(2)}(t))$ is a coupling of two Markov processes: viewing them
separately, $x^{(2)}(t)$ is the original dynamics we are interested in, 
whereas  $x^{(1)}(t)$ is the dynamics on $\Ball_i(r)$ 
(cf. Sec.~\ref{sec:DynamicsDefinition} for a definition) 
with boundary condition $x_{\sim i,r}(0)$.

Consider now a function $f(x_i)$ with $|f| \le 1$. Conditioning on 
$x_{\sim i,r}(0)$, we can apply the observation stated in 
Sec.~\ref{sec:LengthTime}
on the dynamics of $x^{(1)}_{\Ball_i(r)}(t)$ to obtain
\begin{equation}
\widehat{f}(x_{\sim i,r}(0))^2 \le
\< f(x_i^{(1)}(0)) f(x_i^{(1)}(t)) \>_{i,r}^{x(0)} \ . 
\end{equation}
Averaging over $x_{\sim i,r}(0)$, this yields
\begin{equation}
\< f(x_i(0)) \widehat{f}(x_{\sim i,r}(0)) \>  \le
\< f(x_i^{(1)}(0)) f(x_i^{(1)}(t)) \> \ .
\end{equation}
We now introduce the indicator function
\begin{equation}
\ind(t) = \begin{cases} 1 & \mbox{if} \ x_i^{(1)}(t) = x_i^{(2)}(t) \\
0 & \mbox{otherwise}
\end{cases} \ ,
\end{equation}
in terms of which we can rewrite the above inequality as
\begin{eqnarray}
\< f(x_i) \widehat{f}(x_{\sim i,r}) \>   &\le&
\< f(x_i^{(2)}(0)) f(x_i^{(2)}(t))\; \ind (t) \> +
\< f(x_i^{(1)}(0)) f(x_i^{(1)}(t))\; (1-\ind (t)) \> \nonumber \\ &\le&
\< f(x_i^{(2)}(0)) f(x_i^{(2)}(t)) \> + 2 \<1-\ind (t) \> \ . 
\label{eq_lb_in1}
\end{eqnarray}
In the last step we used the fact that $f$ is bounded in absolute value by 1.

A disagreement percolation argument implies the bound
\begin{equation}
\< 1 -\ind(t) \> \le \left( \frac{e \Delta^2 t}{r+1} \right)^{r+1} \ .
\label{eq:disagree}
\end{equation}
Indeed,
$\< 1 -\ind(t) \>$ is the probability that $x^{(1)}$ and $x^{(2)}$ disagree
on the variable node $i$ at time $t$. The coupling between $x^{(1)}(t)$ and
$x^{(2)}(t)$ has been defined in such a way that at initial time the two 
configurations agree on all variable nodes, and that a disagreement on a node
$j \in \Ball_i(r)$ can appear when $j$ is updated only if at least one of
the neighbors of $j$ already bears a disagreeing assignment. In other words,
disagreement has to appear in $V \setminus \Ball_i(r)$ and propagates towards
$i$. More formally, let us call a disagreement path $\alpha$ an ordered list 
of variable nodes $\alpha=(i_r,\dots,i_1,i_0=i)$ such that two successive nodes
are adjacent in $G$, and with $d_{i i_k}=k$. A path $\alpha$ is said to
percolate if there exists a sequence of times $0 < t_1 <\dots < t_{r+1}<t$
such that the clock of the node $i_k$ rings at time $t_{r+1-k}$. The 
quantity
$\< 1 -\ind(t) \>$ is thus bounded by the probability of the event ``at
least one disagreement path has percolated'', which is itself smaller than
the product of the number of such paths by the probability
$p_{r+1}(t)$ for a given path to percolate.
As a given clock rings in the infinitesimal time interval $[t_k,t_k+dt]$ for 
the first time since $t_{k-1}$ with probability $e^{-(t_k-t_k-1)} dt$,
\begin{eqnarray}
p_{r+1}(t) &=& \int_0^t dt_1 \, e^{-t_1} \int_{t_1}^t dt_2 \, e^{-(t_2-t_1)} 
\dots \int_{t_r}^t dt_{r+1} \, e^{-(t_{r+1}-t_r)} = \sum_{s=r+1}^\infty 
\frac{e^{-t} t^s}{s!} \\ 
&\le& \frac{t^{r+1}}{(r+1)!} \le \left( \frac{et}{r+1} \right)^{r+1} \ .
\end{eqnarray}
The number of disagreement paths can be  bounded by $\Delta^{2(r+1)}$,
hence Eq.~(\ref{eq:disagree}).

Subtracting $\< f(x_i) \>^2$ from both sides of Eq.~(\ref{eq_lb_in1}), and
taking the supremum over $|f|\le 1$, we obtain
\begin{equation}
G_i^{(3)}(r) \le C_i(t) + 2 \left( \frac{e \Delta^2 t}{r+1} \right)^{r+1} \ .
\end{equation}
Setting $t=\tau_i(\ve)$ and calling 
$r_* =\max ( 2 e \Delta^2 \tau_i(\ve) ,\,\log_2(2/\ve) )$, we have
\begin{equation}
G_i^{(3)}(r) \le 2 \ve \quad\quad \forall r \ge r_* \ ,
\end{equation}
hence an upperbound on $\ell_i^{(3)}(2 \ve)$, which can be translated into
the form stated in Theorem~\ref{thm:Main} using the inequalities of
Proposition~\ref{propo:CorrEquiv}.
\endproof
%
%
\section{Proof of the upper bound in Theorem \ref{thm:Main}}
\label{sec:ProofUpper}

The upper bound is proved by viewing the dynamics inside
$\Ball_i(r)$ as the dynamics of a `reduced' model whose degrees of freedom
are only the ones in $\Ball_i(r)$, and on which the exterior acts as
a time-dependent boundary condition. On one hand, time correlations
inside the system decay in a finite time (since the system is finite).
On the other, if $r$ is large enough, the time-dependent boundary condition 
does not affect the behavior in the center of $\Ball_i(r)$.

Let us begin by defining the Markov dynamics on a subset of the vertices
$U\subseteq V$ with time dependent boundary conditions $\{y(t)\}_{t\ge 0}$. 
This means that one is given a sequence
of times $t_0=0<t_1<\dots<t_n<\dots$ with $t_n\to\infty$ as $n\to\infty$,
and of configurations $y_0$, $y_1$, $\dots$. The chain is initialized in
a configuration $x$ distributed according to $\mu_U^{y_0}$.
In each  time interval $[t_n,t_{n+1})$, $n=0,1,\dots$,
one runs the chain with boundary condition $y_n$. Then, at time $t_{n+1}$,
the  configuration outside $U$ is changed from $y_n$ to $y_{n+1}$.
Averages with respect to this process will be denoted by 
$\<\, \cdot\,\>^{\{y\}}_U$.

It is convenient to state separately the following estimate on time decay of 
correlations for this dynamics (for the proof, see 
App.~\ref{app:TimeDependent}).
\begin{lemma}\label{lemma:TimeDependent}
Let $f$ and $g$ be two functions of $x\in\cX^N$, such that 
$|f(x)|, |g(x)|\le 1$ for any $x$. Then
\begin{eqnarray}
\Big|\<f(x(0)) g(x(t))\>^{\{y\}}_U-
\<f(x(0))\>^{\{y\}}_U\<g(x(t))\>^{\{y\}}_U\Big|
\le 2\,e^{-(t-1)/\tau_U^*}\, ,
\end{eqnarray}
where $\tau_U^* = \exp\{A|U|\}$, and $A= -\log(\kappa_0(1-e^{-1}))$. 
\end{lemma}

Let us now turn to the actual proof. Fix a vertex $i$, a positive integer 
$r$, and consider a function $f$ of $x_i$, with $|f(x_i)|\le 1$ for all 
$x_i$'s. By conditioning, we can write
\begin{eqnarray}
\<f(x_i(0))f(x_i(t))\> = \E_{\{y\}}\left[\<f(x_i(0))f(x_i(t))\>_{i,r}^{\{y\}}
\right]\, ,
\end{eqnarray}
where $\E_{\{y\}}$ denotes expectation with respect to the process 
$\{y(t)\}_{t\ge 0}$ distributed according to the (stationary)
Markov chain on $G$, and we used the shorthand $\<\,\cdot\,\>^{\{y\}}_{i,r}$
for  $\<\,\cdot\,\>^{\{y\}}_{\Ball_i(r)}$.
Lemma \ref{lemma:TimeDependent} implies
\begin{eqnarray}
\<f(x_i(0))f(x_i(t))\> \le \E_{\{y\}}\left[\<f(x_i(0))\>_{i,r}^{\{y\}}
\<f(x_i(t))\>_{i,r}^{\{y\}}\right] + 2\,e^{-(t-1)/\tau^*(i,r)}\,
\, ,\label{eq:BoundTimeCorrelation}
\end{eqnarray}
where $\tau^*(i,r)$ is a shorthand for $\tau_{\Ball_i(r)}^*$.
The expectation on the right hand side can be simplified by conditioning
on the initial condition $y(0)$:
\begin{eqnarray}
\E_{\{y\}}\left[\<f(x_i(0))\>_{i,r}^{\{y\}}
\<f(x_i(t))\>_{i,r}^{\{y\}}\right] = \E_{y(0)}\left\{
\E_{\{y(t>0)\}}\left[\<f(x_i(0))\>_{i,r}^{\{y\}}
\<f(x_i(t))\>_{i,r}^{\{y\}}\big| \, y(0)\right] 
\right\}=\\
= \E_{y(0)}\left\{ \<f(x_i(0))\>_{i,r}^{y(0)}
\E_{\{y(t>0)\}}\left[\<f(x_i(t))\>_{i,r}^{\{y\}}\big|\, y(0)\right] 
\right\} =\\
= \E_{y(0)}\left[ f(y_i(0)) F(y_{\sim i,r}(0)) \right]
\, , \label{eq:LastCorr}
\end{eqnarray}
where we defined
\begin{equation}
F(y_{\sim i,r}(0)) = \E_{y(0)} \left[
\E_{\{y(t>0)\}}
\left[\<f(x_i(t))\>_{i,r}^{\{y\}} \big| \, y(0)
\right]
\Big| \, y_{\sim i,r}(0) \right] \ .
\end{equation}
Since $f$ is uniformly bounded by $1$, $|F(x_{\sim i,r})|\le 1$
for all $x$ as well. Moreover one easily shows that 
$\<F(x_{\sim i,r})\> = \< f(x_i) \>$.
Subtracting $\< f(x_i) \>^2$ from Eq.~(\ref{eq:BoundTimeCorrelation})
and using the definition of the
spatial correlation function, cf. Eq.~(\ref{eq:SpatialCorrDef}), we obtain
\begin{equation}
C_i^f(t) \le G_i(r) + 2\,e^{-(t-1)/\tau^*(i,r)} \ .
\end{equation}
Taking the supremum over $f$ and setting $r=\ell_i(\ve /2)$ implies
the upper bound of Theorem \ref{thm:Main}.
\endproof
%
%
\section*{Acknowledgments}

A.M. would like to thank Elchanan Mossel for his careful reading of the 
manuscript and a stimulating mail exchange.
A.M. is also grateful to the Newton Institute for 
Mathematical Sciences (Cambridge, UK) for its kind hospitality
during the completion of this work.
G.S. is supported by EVERGROW, integrated project No. 1935 in the
complex systems initiative of the Future and Emerging Technologies directorate 
of the IST Priority, EU Sixth Framework.

%
%
\appendix
%
%
\section{Proof of Proposition \ref{propo:CorrEquiv}}
\label{app:CorrEquiv}

In this appendix we denote simply by $x$ the variable $x_i$, by $y$
the `far apart' variables $x_{\sim i,r}$ and by $\mu(x,y)$ 
(respectively $\mu(x)$, $\mu(y)$) their joint distribution
(respectively, marginal distributions). Finally, we omit the
arguments $i$ and $r$ from the correlation functions
$G_i(r)$, $G^{(1)}_i(r)$, \dots  $G^{(4)}_i(r)$. 
Proving  Proposition \ref{propo:CorrEquiv} amounts to deriving the following
inequalities between these functions:
\begin{align}
\frac{1}{|\cX|}G^{(2)} &\le G\;\;\le  G^{(2)}\, , \label{eq:G}\\
G^{(2)}& \le G^{(1)}\le \frac{2}{\ps}\, G^{(2)}\, ,\label{eq:G1}\\
\left(\frac{1}{|\cX|}G^{(2)}\right)^2&\le G^{(3)}\le G\, ,\label{eq:G3}\\
\frac{1}{2}\, {G^{(2)}}^2 &\le G^{(4)}\le\left(\frac{1}{\ps}-1\right)
G^{(2)}\, .\label{eq:G4}
\end{align}
Proofs of similar statements can be found repeatedly in the literature.
We refer in particular to \cite{GibbsSu} for a general presentation, 
and to \cite{MosselReco} that deals with the tree reconstruction problem 
which is closely related to the theme of this paper. We 
collect nevertheless the proofs here for the sake 
of self-containedness. Notice that the upper 
bounds in Eqs.~(\ref{eq:G1}) and (\ref{eq:G4}) become trivial if
$\mu_*=0$. We shall therefore assume, without loss of generality,
$\mu_*>0$.

\underline{(\ref{eq:G}), lower bound.}
In the definition of $G$, take~\footnote{We denote by $\ind(A)$
the indicator function for the property $A$.} $f(x) = \ind(x=x_*)$, and
$F(y) = \sign[\mu(y|x_*)-\mu(y)]$. Then 
\begin{eqnarray}
G \ge \sum_{x,y}f(x)F(y)\mu(x)[\mu(y|x)-\mu(y)]= \mu(x_*)
\sum_y\left|\mu(y|x_*)-\mu(y)\right|\, .
\end{eqnarray}
The thesis follows by choosing $x_*$ which maximizes the last expression.

\underline{(\ref{eq:G}), upper bound.} We have
\begin{eqnarray}
G = \sup_{f,F}\left|\sum_{x,y}f(x)F(y)[\mu(x,y)-\mu(x)\mu(y)]\right|
\le \sum_{x,y}\left|[\mu(x,y)-\mu(x)\mu(y)]\right|\,,
\end{eqnarray}
that is what was claimed.

\underline{(\ref{eq:G1}), lower bound.} We have
\begin{eqnarray}
G^{(2)} = \sum_{x,y}\mu(x)\left|\sum_{x'}\mu(x')[\mu(y|x)-\mu(y|x')]
\right|\le \sum_{x,x'}\mu(x)\mu(x')\sum_y
\left|\mu(y|x)-\mu(y|x')\right|\, ,
\end{eqnarray}
And the last expression is upper bounded as 
$\sup_{x,x'}\sum_y \left|\mu(y|x)-\mu(y|x')\right|\equiv G^{(1)}$.

\underline{(\ref{eq:G1}), upper bound.} We start by noticing that 
\begin{eqnarray}
G^{(2)} =\sum_x \mu(x)\sum_y \left|\mu(y|x)-\mu(y)\right|
\ge \mu_*\, \sup_x\l|\mu_{Y|X}(\,\cdot\,|x)-\mu_Y(\,\cdot\,)\r|_1\, ,
\end{eqnarray}
where we introduced the standard notation for $L^1$ norm 
and used subscripts to precise which variable we are considering the
distribution of. By triangular inequality 
$||\mu_{Y|X}(\,\cdot\,|x_1)-\mu_{Y|X}(\,\cdot\,|x_2)||_1\le
2\, \sup_x||\mu_{Y|X}(\,\cdot\,|x)-\mu_Y(\,\cdot\,)||_1$
for any $x_1$, $x_2$. The thesis follows by taking the $\sup$
over $x_1$, $x_2$.

\underline{(\ref{eq:G3}), lower bound.}
Take $f(x) =\ind(x=x_*)$, and therefore $\fp(y) = \mu(x_*|y)$.
Then
\begin{eqnarray}
G^{(3)} \ge \sum_{x,y}f(x)\fp(y)[\mu(x,y)-\mu(x)\mu(y)] = 
\sum_y\frac{1}{\mu(y)}[\mu(x_*,y)-\mu(x_*)\mu(y)]^2\, .
\end{eqnarray}
By maximizing the right hand side over $x_*$, we obtain
\begin{eqnarray}
G^{(3)} & \ge &\frac{1}{|\cX|}
\sum_{x,y}\frac{1}{\mu(y)}[\mu(x,y)-\mu(x)\mu(y)]^2=\\
&= & \frac{1}{|\cX|^2}
 \left(\sum_{x,y}\frac{1}{\mu(y)}\mu(y)^2\right)
\left(\sum_{x,y}\frac{1}{\mu(y)}[\mu(x,y)-\mu(x)\mu(y)]^2\right)\ge
\nonumber\\
&\ge &\frac{1}{|\cX|^2}
\left(\sum_{x,y}|\mu(x,y)-\mu(x)\mu(y)|\right)^2
\, ,
\end{eqnarray}
where the last step followed from Cauchy-Schwarz inequality.

\underline{(\ref{eq:G3}), upper bound.} Trivial: take $F=\fp$.

\underline{(\ref{eq:G4}), lower bound.} We notice that 
$I(X;Y) = D(p||q)$ where $p$ and $q$ are distributions on the pair
$z=(x,y)$ defined by $p(x,y)=\mu(x,y)$ and $q(x,y)=\mu(x)\mu(y)$,
and $D(p||q) = \sum_zp(z)\log[p(z)/q(z)]$. Defining
$p_{\lambda}(z) = (1-\lambda)q(z)+\lambda p(z)$, by elementary calculus
\begin{eqnarray}
D(p||q) = \int_0^1(1-\lambda)
\sum_z\frac{1}{p_{\lambda}(z)}[p(z)-q(z)]^2\;\de\lambda\, .\label{eq:KLRepres}
\end{eqnarray}
By Cauchy-Schwarz (applied to the scalar product with weight
$1/p_{\lambda}(z)$) we have $\sum_z\frac{1}{p_{\lambda}(z)}[p(z)-q(z)]^2\ge
(\sum_z|p(z)-q(z)|)^2$, and the thesis follows.

\underline{(\ref{eq:G4}), upper bound.}
We use $p_{\lambda}(z)\ge(1-\lambda)q(z)$ in Eq.~(\ref{eq:KLRepres}),
and get
\begin{eqnarray}
G^{(4)}\le\sum_{z} \frac{1}{q(z)}[p(z)-q(z)]^2\le
\sup_{z}\left|\frac{p(z)}{q(z)}-1\right|\sum_{z}|p(z)-q(z)|\, .
\end{eqnarray}
The thesis follows by noticing that $p(z)/q(z)= \mu(x|y)/\mu(x)\le 1/\mu_*$.
%
%
\section{Proof of Lemma \ref{lemma:TimeDependent}}
\label{app:TimeDependent}

Lemma \ref{lemma:TimeDependent} is a well known elementary result for
Markov chains with time-independent boundary conditions
(see for instance \cite{Guionnet} for a functional-analytic argument).
We present here an independent and self-contained proof for the general 
case. We start by restating it in a
slightly stronger form. For this purpose, we need to define the 
dynamics on $U\subseteq V$, with time-dependent boundary condition
$\{y(t),\, t\ge 0\}$ and generic initial distribution 
$\nu$. This is defined exactly as the process with time-dependent 
boundary condition introduced in Sec.~\ref{sec:ProofUpper}, 
but for the fact that the
initial state is distributed according to $\nu$, instead of 
$\mu^{y(0)}_U$. It is understood that $\nu(x)=0$ unless 
$x_{V\backslash U}  = y(0)_{V\backslash U}$. 
\begin{lemma}\label{lemma:TimeDepDistance}
Let $\nu^{(1)}_t$ and $\nu^{(2)}_t$ the distributions at time $t$ for the 
dynamics on $U\subseteq V$, with the same time-dependent boundary condition
$\{y(t),\, t\ge 0\}$, and initial distributions (respectively),
$\nu^{(1)}$ and $\nu^{(2)}$. Then
\begin{eqnarray}
||\nu^{(1)}_t-\nu^{(2)}_t||_\sTV\le e^{-(t-1)/\tau_U^*}
||\nu^{(1)}-\nu^{(2)}||_\sTV\, ,
\end{eqnarray}
where $\tau_U^* = \exp\{A|U|\}$, and $A= -\log(\kappa_0(1-e^{-1}))$.
\end{lemma}
\prooft
Denote by $\{x^{(\alpha)}(t), t\ge 0\}$, $\alpha\in\{1,2\}$ the two processes.
We construct a coupling of these two processes, similar to the one
of Sec.~\ref{sec:ProofLower}:
\begin{itemize}
\item[$\bullet$] at the initial time $t=0$, $x^{(1)}(0)$ 
and $x^{(2)}(0)$ are drawn from the greedy coupling of $\nu^{(1)}$ and
$\nu^{(2)}$, hence 
$\prob\left[x^{(1)}(0)\neq x^{(2)}(0)\right]=||\nu^{(1)}-\nu^{(2)}||_\sTV$.
\item[$\bullet$] When the clock at $j\in U$ rings, say at time $t$, 
we draw $(\xi_1,\xi_2)$ from the greedy coupling
between $\kappa_j^{x^{(1)}(t)}$ and $\kappa_j^{x^{(2)}(t)}$, and replace
$(x_j^{(1)}(t),x_j^{(2)}(t))$ by $(\xi_1,\xi_2)$.
\end{itemize}

Consider now two times $t$ and $t'=t+\Delta t>t$. 
If the two processes coincide 
at a given time, the definition of the coupling implies that they will
coincide at all subsequent times. Therefore
\begin{eqnarray}
\prob\left[ x^{(1)}(t')\neq x^{(2)}(t')\right]
=\left\{
1-\prob\left[ x^{(1)}(t')= x^{(2)}(t')\big| 
x^{(1)}(t)\neq x^{(2)}(t)\right]\right\}\, 
\prob\left[ x^{(1)}(t)\neq x^{(2)}(t)\right]
\, .\label{eq:OneInterval}
\end{eqnarray}
The conditional probability appearing in the last expression
can be lower bounded by the probability of a particular event 
implying $x^{(1)}(t')= x^{(2)}(t')$ irrespective of
$x^{(1)}(t)$, $x^{(2)}(t)$. The event is defined as follows.
Each variable in $U$ tries at least one flip during the time
interval $[t,t')$ (this happens with probability $(1-e^{-\Delta t})^{|U|}$).
Furthermore, the last time a flip is attempted on each of the 
spins, it brings the two coupled processes to coincide
on it (this happens with probability at least $\kappa_0$ for each spin).
We have therefore
\begin{eqnarray}
\prob\left[ x^{(1)}(t')= x^{(2)}(t')\big| 
x^{(1)}(t)\neq x^{(2)}(t)\right]\ge 
\left[\kappa_0(1-e^{-\Delta t})\right]^{|U|}\, .
\end{eqnarray}
Finally consider the time interval $[0,t]$ and split it
into $\lfloor t/\Delta t\rfloor$ sub-intervals of size $\Delta t$
(plus, eventually a sub-interval of smaller size).
By repeatedly applying Eq.~(\ref{eq:OneInterval}), recalling
that $(1-x)\le e^{-x}$ and $||\nu^{(1)}_t-\nu^{(2)}_t||_\sTV \le
\prob\left[ x^{(1)}(t) \neq x^{(2)}(t)\right]$, we get
\begin{eqnarray}
||\nu^{(1)}_t-\nu^{(2)}_t||_\sTV\le
\exp\left\{-\left[\kappa_0(1-e^{-\Delta t})\right]^{|U|} 
\left\lfloor\frac{t}{\Delta t}\right\rfloor\right\}
||\nu^{(1)}-\nu^{(2)}||_\sTV
\end{eqnarray}
The thesis is proved by taking $\Delta t=1$, and noticing that
$\lfloor t\rfloor\ge t-1$.
\endproof

We next show that this result implies Lemma \ref{lemma:TimeDependent}.
We have
\begin{eqnarray}
\Big|\<f(x(0)) g(x(t))\>^{\{y\}}_U-
\<f(x(0))\>^{\{y\}}_U\<g(x(t))\>^{\{y\}}_U\Big| =\\
=\left|\sum_{x,x'}\prob\{x(0)=x\}f(x)g(x')\left[
\prob\left\{x(t)=x'\big|x(0)=x\right\}-\prob\left\{x(t)=x'\right\}
\right]\right|\le\\
\le\sum_{x}\prob\{x(0)=x\}\sum_{x'}
\left|\prob\left\{x(t)=x'\big|x(0)=x\right\}-\prob\left\{x(t)=x'\right\}\right|
\, .
\end{eqnarray}
The sum over $x'$ in the last expression is 
$2||\nu^{(1)}_t-\nu^{(2)}_t ||_\sTV$, for two processes of initial conditions
$\nu^{(1)}(x')=\ind(x=x')$ and $\nu^{(2)}=\mu_U^{y(0)}$. The proof is 
completed by applying Lemma \ref{lemma:TimeDepDistance}, with
$||\nu^{(1)}-\nu^{(2)} ||_\sTV \le 1$.
\endproof
%
%
\section{High-temperature upper bound for the $p$-spin model}
\label{sec:HighTPspin}

In this Appendix we prove the first part of Proposition \ref{propo:TimePspin}:
at high enough temperature, the correlation time $\tau_i(\ve)$ is,
with high probability, finite.

We begin by recalling that the temperature $\beta^{\rm fast}_{p,l}=1/
T_{p,l}^{\rm fast}$ appearing in the statement of
Proposition \ref{propo:TimePspin} is defined as the largest value of 
$\beta$ such that
\begin{eqnarray}
(p-1)\,l\,\tanh\beta \le 1\, .
\end{eqnarray}
We shall also use the notation $\rho(\sigma,\tau)$ to denote the Hamming 
distance (number of different spins) between two configurations 
$\sigma$ and $\tau$. 
The proof makes use of the following crucial result.
\begin{lemma}
Consider the $p$-spin model, cf. 
Eqs.~(\ref{eq:PspinBoltzmann}) and (\ref{eq:PspinEnergy}), at inverse 
temperature $\beta<\beta^{\rm fast}_{p,l}$, and 
let $\mu^{(i)}_{\pm}$ be  
the Boltzmann measure, conditioned to $\sigma_i=\pm 1$. 
Then  there exists a coupling of $\mu^{(i)}_+$ and
 $\mu^{(i)}_{-}$ such that, 
if $(\sigma,\tau)$ is a pair of configurations distributed according to such a
coupling, their expected Hamming distance is bounded as
\begin{eqnarray}
\<\rho(\sigma,\tau)\>\le \frac{1}{1-l(p-1)\tanh\beta} \, . 
\label{eq:BoundHamming}
\end{eqnarray}
\end{lemma}
\prooft
Without loss of generality, we set $i=0$. 

Given  a coupling $\nu$ between $\mu^{(0)}_+$ and $\mu^{(0)}_-$, 
and  a pair of configurations $\sigma$, $\tau$ distributed 
according to $\nu$, let $p_j(\nu)$ be the probability that $\sigma_j\neq\tau_j$
under  $\nu$. By definition,  $p_0(\nu)=1$. We claim that, given $j \in  V$, 
$j\neq 0$, it is possible to construct another coupling $\nu'$ 
between $\mu^{(0)}_+$ and $\mu^{(0)}_-$ 
in such a way that 
\begin{equation}
\begin{cases}
p_j(\nu') \le  
\tanh\beta\, \underset{a\in\partial j}{\sum} \ 
\underset{k\in\partial a\backslash j}{\sum}
p_k(\nu)\, ,\label{eq:Claim} \\
p_k(\nu') = p_k(\nu) \ \forall k \neq j
\end{cases} \ .
\end{equation}
We shall denote in the following $\T_j$ the mapping $\nu' = \T_j \nu$.
The coupling $\nu'$ can be defined through  
the following sampling procedure.
First sample $\sigma$ and $\tau$ from $\nu$. Then draw $\sigma'_j$, $\tau'_j$ 
by coupling in a greedy fashion the conditional distributions 
$\mu(\sigma_j |\sigma_{\sim j})$, and $\mu(\tau_j|\tau_{\sim j})$
(to be explicit, we shall denote these conditional distributions
as $\mu_j(\,\cdot\,|\sigma_{\sim j})$, and $\mu_j(\,\cdot\,|\tau_{\sim j})$
in the following). 
Finally set $\sigma'_k = \sigma_k$ and $\tau'_k=\tau_k$
for all nodes $k\neq j$, and define $\nu'$ to be the joint
distribution of $\sigma'$ and $\tau'$.
Obviously $p_k(\nu')= p_k(\nu)$ for all $k\neq j$. Moreover
\begin{eqnarray}
p_j(\nu') = \sum_{\sigma,\tau}\nu(\sigma,\tau)\,
||\mu_j(\,\cdot\, |\sigma_{\sim j})-
\mu_j(\, \cdot\, |\tau_{\sim j})||_{\sTV}\, .
\end{eqnarray} 
Notice that $\mu_j(\,\cdot\, |\sigma_{\sim j})$ depends on 
$\sigma_{\sim j}$ only through $\sigma_a \equiv \{\sigma_{k}:\,
k\in\partial a\backslash j\}$, with $a\in \partial j$. Denote by  
$a_1,\dots a_l$ the indices\footnote{Within the configuration model
is possible (although with probability vanishing as $N\to\infty$)
that the variable node $j$ has less than $l$ neighboring function nodes.
Although the proof remains valid in this case, we shall not consider it
explicitly in order to lighten the notation.} 
of function nodes which are neighbors of $j$
and define $\sigma^{(t)}$, $t=0,\dots, l$ in such a way that 
$\sigma^{(t)}_{a_s} = \sigma_{a_s}$ for $s\le t$ and 
$\sigma^{(t)}_{a_s} = \tau_{a_s}$  for $s> t$ (in particular
$\sigma^{(0)} = \tau$ and $\sigma^{(l)} = \sigma$). Then
\begin{eqnarray}
p_j(\nu') & \le &\sum_{\sigma,\tau}\nu(\sigma,\tau)\,\sum_{t=1}^{l}
||\mu_j(\,\cdot\, |\sigma^{(t)}_{\sim j})-
\mu_j(\, \cdot\, |\sigma^{(t-1)}_{\sim j})||_{\sTV}\le\\
&\le & \sum_{\sigma,\tau}\nu(\sigma,\tau)\,\sum_{t=1}^{l}
\ind(\sigma_{a_t}\neq \tau_{a_t})\tanh\beta =
\tanh\beta \sum_{t=1}^{l}\prob_{\nu }(\sigma_{a_t}\neq \tau_{a_t})
\, .\label{eq:BoundNeighbors}
\end{eqnarray} 
Here we used the fact that 
\begin{eqnarray}
||\mu_j(\,\cdot\, |\sigma^{(t)}_{\sim j})-
\mu_j(\, \cdot\, |\sigma^{(t-1)}_{\sim j})||_{\sTV}=\frac{1}{2}\left|
\tanh\left(\beta \sum_{a\in\partial j }J_a\!\!\prod_{k\in\partial a\backslash j}\sigma^{(t)}_k\right)-\tanh\left(\beta \sum_{a\in\partial j }J_a\!\!\prod_{k\in\partial a\backslash j}\sigma^{(t-1)}_k\right)
\right|\, ,
\end{eqnarray}
and $|\tanh a-\tanh b|\le 2\tanh(|a-b|/2)$. Applying the union bound
to Eq.~(\ref{eq:BoundNeighbors}), we get Eq.~(\ref{eq:Claim}).

We shall now construct another mapping $\nu' = \T \nu$ by combining
the elementary $\T_j$. Defining $\nu^{(0)} = \nu$, and
ordering arbitrarily the variable nodes $1, \dots, N-1$,
we construct recursively the couplings $\nu^{(1)},\dots$,
$\nu^{(N-1)}$  with
\begin{equation}
\nu^{(j)} =
\begin{cases}
\T_j \nu^{(j-1)} & {\rm if} \ p_j(\T_j \nu^{(j-1)}) \le p_j(\nu^{(j-1)}) \\
\nu^{(j-1)} & {\rm otherwise}
\end{cases} \quad , \ {\rm for} \ j=1,\dots,N-1 \ .
\end{equation}
Finally we let $\nu^{(N-1)}\equiv \T\nu$. A moment of thought shows that
\begin{eqnarray}
p_j(\T\nu)\le 
\tanh\beta\, \sum_{a\in\partial j}\sum_{k\in\partial a\backslash j}
p_k(\nu)\, ,\label{eq:Tcoupl}
\end{eqnarray}
for all $j\neq 0$.

To conclude the proof, denote by $\orh(\nu)$ the expectation of
$\rho(\sigma,\tau)$ when $\sigma$ and $\tau$ are distributed according
to the coupling $\nu$, which can also be rewritten as 
$\orh(\nu) =\sum_{j}p_j(\nu)$.
By summing over $j\neq 0$ Eq.~(\ref{eq:Tcoupl}), recalling that 
$p_0(\nu) =1$ and that each node $k$ is the neighbor of at most
$l(p-1)$ nodes $j$, we get
\begin{eqnarray}
\orh(\T\nu)\le 1+l(p-1)\,\tanh\beta\, \orh(\nu)\,. 
\end{eqnarray}
Since $l(p-1)\,\tanh\beta < 1$, a coupling achieving
(\ref{eq:BoundHamming}) can be constructed by iterating a sufficient number 
of times the transformation $\nu\mapsto\T\nu$ from an arbitrary initial 
coupling. In fact the sequence of distributions $\T^n\nu$
admits a subsequential limit because the space of distributions over 
a finite set is compact.
\endproof

Notice that the above proof is in fact closely related to the proof of 
Dobrushin uniqueness condition~\cite{Dobrushin} as described, for instance, in 
\cite{Georgii}. For arguments of this type we also refer to
\cite{Weitz}.

We are now in position of proving the upper bound in 
Proposition \ref{propo:TimePspin}.
Consider two processes $\{\sigma^{(+)}(t)\}_{t\ge 0}$
and $\{\sigma^{(-)}(t)\}_{t\ge 0}$ evolving according to the Glauber 
dynamics with initial conditions distributed accordingly 
to $\mu_+^{(i)}$ (for $\sigma^{(+)}(0)$),  and $\mu_-^{(i)}$
(for $\sigma^{(-)}(0)$). In other words, $\{\sigma^{(+)}(t)\}$ 
(respectively $\{\sigma^{(-)}(t)\}$) is the stationary Glauber dynamics
conditioned to $\sigma^{(+)}_i(0) = +1$ (respectively, 
$\sigma^{(-)}_i(0) = -1$).  Then it is easy to show that
\begin{eqnarray}
C_i(t) = \frac{1}{2}(1-m_i^2)\left[\<\sigma^{(+)}_i(t)\>-\<\sigma_i^{(-)}(t)\>
\right]\, ,
\end{eqnarray}
where $m_i= \<\sigma_i\>$ is the expectation of $\sigma_i$ with respect to the
(unconditional) Boltzmann measure.

Given an arbitrary coupling of the two processes $\{\sigma^{(+)}(t)\}$,
and  $\{\sigma^{(-)}(t)\}$, the correlation function is obviously bounded as
\begin{eqnarray}
C_i(t)\le \, \left\<\,\rho\left(\sigma^{(+)}(t),\sigma^{(-)}(t)
\right)\right\>\, .
\end{eqnarray}

We construct such a coupling as follows. The initial conditions 
$\sigma^{(+)}(0)$, and $\sigma^{(-)}(0)$ are chosen according to 
a coupling of the conditional distributions $\mu^{(i)}_{\pm}$
that achieves the bound (\ref{eq:BoundHamming}). The joint dynamics
is defined using `path coupling' \cite{BubleyDyer,AldousFill}. 
This construction only requires to define 
the  evolution of $\sigma^{(+)}(t)$ and $\sigma^{(-)}(t)$ when they 
differ in a single position. If this is the case, we use the greedy coupling
of the update probabilities, as in the proof of 
Lemma \ref{lemma:TimeDependent}, cf. App.~\ref{app:TimeDependent}. 
From a pair of configurations $(\sigma^{(+)},\sigma^{(-)})$ with
$\rho(\sigma^{(+)},\sigma^{(-)})=1$, the expected (with respect to the greedy 
coupling) Hamming distance after one spin update can be upper bounded by
$1-(\kappa/N)$, with $\kappa \equiv (1-l(p-1)\tanh\beta)$. Standard path
coupling arguments allow to extend this bound for an arbitrary number
of spin updates.
Let us denote by $U_t$ the number of spin flips between times $0$
and $t$ (i.e. a Poisson variable of mean $Nt$), and by $\E_U$ the
corresponding expectation:
\begin{eqnarray}
C_i(t)\le \, 
\left\<\,\rho\left(\sigma^{(+)}(0),\sigma^{(-)}(0)\right)\right\> 
\E_U \left(1-\frac{\kappa}{N}\right)^{U_t}
\le \kappa^{-1} \, \exp\left\{-\kappa t\right\}
\, ,
\end{eqnarray}
which clearly implies the thesis.
\endproof
%
%
\section{Low-temperature lower bound for the $p$-spin model}
\label{sec:LowTPspin}

In this Appendix we prove the second part of 
Proposition \ref{propo:TimePspin} on the
correlation time of the $p$-spin model on random regular hypergraphs.
Throughout the Appendix we denote by 
$Q_{\sigma,\tau}=N^{-1}\sum_{i=1}^N\sigma_i\tau_i$ the normalized overlap
of configurations $\sigma$ and $\tau$. We further let 
$Z(\beta)$ be the partition function, i.e. the
normalization constant in Eq.~(\ref{eq:PspinBoltzmann}),
and $Z(q;\beta)$ the {\em constrained} partition function
\begin{eqnarray}
Z(q;\beta)\equiv\sum_{\sigma^{(1)},\sigma^{(2)}}
e^{-\beta E(\sigma^{(1)})-\beta E(\sigma^{(2)})}\,
\ind\left(Q_{\sigma^{(1)},\sigma{(2)}} = q\right)\, ,
\end{eqnarray}
where $q\in\{-1,-1+2/N,\dots,1-2/N,1\}$. 

It is convenient to state a few preliminary results.
We start by computing the expectation of $Z(\beta)$ and
$Z(\beta,q)$. It is straightforward to get
\begin{eqnarray}
\E\, Z(\beta) = 2^{N}(\cosh\beta)^M = e^{N\phi(\beta)}
\end{eqnarray}
where we defined $\phi(\beta) = \log 2+\frac{l}{p}\log\cosh\beta$.
The expected  constrained partition function is only slightly more involved

\begin{eqnarray}
\E\, Z(q;\beta) = 2^N\binom{N}{m}\, \binom{Nl}{ml}^{-1}
\coeff[((\cosh 2\beta) p_+(x) + p_-(x))^M,x^{(N-m)l}] 
\doteq e^{N\phi(q;\beta)}
\end{eqnarray}
where we denoted by $\doteq$ identity to the leading 
exponential order, $m\equiv N(1+q)/2$, and $p_{\pm}(x) = 
\frac{(1+x)^p\pm (1-x)^p}{2}$. Using Haymann (or saddle-point) method, one
obtains the exponential growth rate 
\begin{eqnarray}
\phi(q;\beta) & = & \log 2  -(l-1)\, h\!\left(\frac{1-q}{2}\right)
+ \frac{l}{p} \log \left( (\cosh 2\beta) p_+(z_q) +  p_-(z_q)  \right)
- l \frac{1-q}{2} \log z_q \ ,
\label{eq:AnnealedPot1}
\end{eqnarray}
where $h(x) = -x\log x-(1-x)\log(1-x)$ is  the entropy function and
$z_q$ is the unique non-negative solution of
the equation
\begin{eqnarray}
z_q \frac{(\cosh 2\beta) p'_+(z_q) +  p'_-(z_q)}
{(\cosh 2\beta) p_+(z_q) +  p_-(z_q)} = p\, \frac{1-q}{2} \ .
\label{eq:AnnealedPot2}
\end{eqnarray}
The function $\phi(q;\beta)$ is straightforwardly evaluated numerically. 

Let us underline
some of its properties. One easily shows that for all temperatures, it has
a local maximum at $q=0$, with $\phi(0;\beta)=2 \phi(\beta)$.
Moreover, for $\beta = 0$, the function reduces to 
$\phi(q;0)=\log 2 + h((1-q)/2)$, for which $q=0$ is the global maximum.
We define $\beta^{\rm ann}_{p,l}$ as the largest value
of $\beta$ such that for any $\beta'<\beta$, $\phi(q;\beta')$ achieves its 
global maximum at $q=0$. Notice  that, if $p\ge l$, 
then $\beta^{\rm ann}_{p,l}=\infty$.

Let us now turn to the behaviour around $q=1$.
A simple calculation shows that $\phi(q=1;\beta) = \phi(2\beta)$.
Let us define $U(q;\beta)\equiv  \phi(q=1;\beta)-\phi(q;\beta)$, 
and $U(q;\infty)$ its limit as $\beta \to \infty$. Taking afterwards the limit
$q\to 1$, one shows that 
$U(1-\delta;\infty) = - (l-2)(\delta \log \delta)/4 + O(\delta)$.

We define the {\em annealed free energy barrier}
\begin{eqnarray}
\Upsilon(\beta) & \equiv & \sup_{\alpha\in (1/2,1)}
[(1-\alpha)\Upsilon_0(\beta) -\ph(\beta,\alpha)]\, ,
\label{eq:BarrierDef}\\
\Upsilon_0(\beta) & \equiv & \sup_{q\in(0,1]} U(q;\beta) \, ,
\label{eq:Barrier0Def}\\
\ph(\beta,\alpha) 
& \equiv & \alpha\phi\left((2-\alpha^{-1})\beta\right)
+(1-\alpha)\phi(2\beta)-\phi(\beta)\, . \label{eq:DefPh}
\end{eqnarray}

From the behaviour of $U(q;\infty)$ in the limit $q \to 1$ one can
deduce that, if $l\ge 3$, $\Upsilon_0(\beta)$ is strictly 
positive for $\beta$ large enough.
On the other hand, $\ph(\beta,\alpha)$ is non-negative because 
$\phi(\beta)$ is convex. From the low temperature expansion
$\phi(\beta) = (1-l/p)\log 2+(l/p)\beta+O(e^{-2\beta})$ it follows
that $\ph(\beta,\alpha)\to 0$ as $\beta\to\infty$.
We proved therefore that, if $l\ge 3$, there exists 
$\beta <\infty $ such that 
$\Upsilon(\beta') >0$ for $\beta'>\beta$. 
We call $\beta_{p,l}^{\rm barr}$ the smallest $\beta$ with such a property.
The numerical estimation of the associated temperatures (inverse of $\beta$) 
for a few values of $(p,l)$ are given in Table~\ref{tab:temperatures}.

These computations have several immediate consequences. The first is that
they yield the free energy in the thermodynamic limit.
\begin{lemma}\label{lemma:PartF}
Consider the model (\ref{eq:PspinBoltzmann}), (\ref{eq:PspinEnergy})
with $p \ge 3$ and $l\ge 2$, and let $\phi(\beta)$ be given as above. 
Define the expected free energy density as 
$\phi_N(\beta) \equiv N^{-1}\E\log Z(\beta)$, and assume
$\beta< \beta^{\rm ann}_{p,l}$.
Then $\phi_N(\beta)\to\phi(\beta)$ as $N\to\infty$.
Furthermore, for any $\delta>0$ there exists
$C(\delta)>0$ such that:
\begin{eqnarray}
\prob\left\{\left|\log Z(\beta)-N\phi_N(\beta)\right|\ge N\delta\right\}
\le \, 2\,e^{-NC(\delta)}\, .\label{eq:Concentration}
\end{eqnarray}
\end{lemma}
\prooft
The statement $\phi_N(\beta)\to\phi(\beta)$ follows from the second moment 
method applied to the random variable $Z(\beta)$ (notice that 
$\E\, Z(\beta)^2 \doteq \exp\{N\sup_q\phi(q;\beta)\}$). Equation
(\ref{eq:Concentration}) can then be proved through
standard concentration inequalities \cite{TalaBook}. 
\endproof

\begin{lemma}\label{lemma:ConstrPartF}
Consider the model (\ref{eq:PspinBoltzmann}), (\ref{eq:PspinEnergy}) with 
$p \ge 3$ and $ l\ge 2$,
and let $\phi(q;\beta)$ be defined as in Eqs.~(\ref{eq:AnnealedPot1}),
(\ref{eq:AnnealedPot2}). Then, 
for any $\beta>0$ and any $\delta>0$, and any $N$ large enough:
\begin{eqnarray}
\prob\left\{Z(q;\beta)\ge e^{N[\phi(q;\beta)+\delta]}\right\}
\le \, e^{-N\delta/2}  \, .
\end{eqnarray}
\end{lemma}
\prooft
This is just Markov inequality applied to the random variable $Z(q;\beta)$,
noting that $\lim (1/N) \log \E Z(q;\beta) = \phi(q;\beta)$.
\endproof

\begin{lemma}\label{lemma:OverlapConcentration}
Consider the model (\ref{eq:PspinEnergy}) with 
$p \ge 3$ and $ l\ge 2$, and $\beta<\beta^{\rm ann}_{p,l}$. Let 
$\sigma^{(1)}$ and $\sigma^{(2)}$ be two i.i.d. configurations
drawn from the Boltzmann distribution (\ref{eq:PspinBoltzmann}),
and denote by $Q_{12}$
be their overlap. Then, for any $\delta>0$, 
there exist constants $C_1$ and $C_2>0$ such that, for all
$N$ large enough 
\begin{eqnarray}
\prob\left\{\left|\<Q_{12}\>\right|\ge N\delta \right\}
\le C_1(\delta)\,  e^{-NC_2(\delta)}\, .
\end{eqnarray}
\end{lemma}
\prooft
First notice that, by the two previous Lemmas, there exist constants
$C_1(\delta)$ and $C_2(\delta)$ such that, with probability
at least $1-C_1(\delta)\, e^{-NC_2(\delta)}$, the following happens:
$(i)$ $Z(q;\beta)\le e^{N[\phi(q;\beta)+\delta^2]}$ for
any $q\in\{-1,-1+2/N,\dots,1-2/N,1\}$, and 
$(ii)$ $Z(\beta)\ge e^{N[\phi(\beta)-\delta^2]}$.
Under these conditions, for any $\xi>0$ we have
\begin{eqnarray}
\<Q_{12}\>&\le &\xi +\prob\left\{|Q_{12}|\ge\xi\right\}=
\xi +\frac{1}{Z(\beta)^2}\sum_{|q|\ge \xi} Z(q;\beta)\\ 
&\le &\xi+e^{3N\delta^2}\sum_{|q|\ge\xi}\exp\left\{
N\left[\phi(q;\beta)-2\phi(\beta)\right]\right\}\, ,
\end{eqnarray}
with the sum being restricted to $q\in\{-1,-1+2/N,\dots,1-2/N,1\}$.
Recall that $q=0$ is a stationary point of $\phi(q;\beta)$
with $\phi(0,\beta)=2\phi(\beta)$. A little 
calculus shows that $\phi''(0;\beta)<0$ for any $p\ge 3$. As a consequence
for any $T>T^{\rm ann}_{p,l}$ there exists $\alpha>0$ such that 
$\phi(q;\beta)\le 2\phi(\beta)-\alpha q^2$.
Therefore
\begin{eqnarray}
|\<Q_{12}\>|\le \xi +2N e^{3N\delta^2-N\alpha\xi^2}\, .
\end{eqnarray}
By taking $\xi = 2\delta/\sqrt{\alpha}$, we have
$|\<Q_{12}\>|\le  4\delta/\sqrt{\alpha}$ for any $N$ large enough.
The thesis follows by rescaling $\delta$.
\endproof

The crucial step is made in the next Lemma, which estimates the global
correlation function 
\begin{eqnarray}
\tC(t) = \frac{1}{N}\sum_{i=1}^N \<\sigma_i(0)\, \sigma_i(t)\>\, .
\end{eqnarray}
\begin{lemma}\label{lemma:Corr}
Consider the model (\ref{eq:PspinBoltzmann}), (\ref{eq:PspinEnergy})
 with $p \ge 3$, $l\ge 2$  and 
$\beta^{\rm barr}_{p,l}\le \beta\le \beta^{\rm ann}_{p,l}$.
Let $\Upsilon= \Upsilon(\beta)$ be the associated  annealed free energy barrier,
and $q_*$ be the largest value of $q$ at which the 
$\sup$ in Eq.~(\ref{eq:Barrier0Def}) is achieved.
Then, for any $\delta\in (0,1/4]$, 
and $t>(4/N)\,\log(2/\delta)$, one has 
$\tC(t)\ge q_*-\delta-t\, e^{-N[\Upsilon-\delta]}$
with high probability.
\end{lemma}
\prooft 
Consider two equilibrium trajectories $\{\sigma^{(1)}(t):\, t\ge 0\}$, 
 $\{\sigma^{(2)}(t):\, t\ge 0\}$ evolving independently according to the
stationary Glauber dynamics for the model 
(\ref{eq:PspinBoltzmann}), (\ref{eq:PspinEnergy}).  In particular,
at any time $\sigma^{(1)}(t)$ and $\sigma^{(2)}(t)$ are distributed 
independently according to the equilibrium distribution $\mu$. 
We further let
$Q(t) \equiv Q_{\sigma^{(1)}(t),\sigma^{(2)}(t)}$.
Throughout the proof, we denote by $\Ep$ and $\Pp$, respectively,
expectation and probability with respect to this process
(not to be confused with expectation and probability with respect to the 
graph and energy function).

It is elementary that 
\begin{eqnarray}
\tC(2t) = \sum_{\sigma}\mu_{\beta}(\sigma)\, 
\Ep \left[Q(t)| \sigma^{(1)}(0)= \sigma^{(2)}(0)=\sigma\right]\, .
\end{eqnarray}
where, for future convenience, we specified the temperature
$\beta$ at which the Boltzmann measure must be considered. 
Let us denote by $I_{\sigma}$ the event 
$\{\sigma^{(1)}(0)= \sigma^{(2)}(0)=\sigma\}$.
Moreover, for any
$q\in\{-1,-1+2/N,\dots, 1-2/N,1\}$, we denote by 
$\Ev_{q,t}$ the event that there exists a time $s\in[0,t]$ such that
$Q(s)=q$. Clearly
\begin{eqnarray}
 \Ep \left[Q(t)| I_{\sigma}\right] &=&  \Pp\{\Ev_{q,t}|I_{\sigma}\}
\Ep \left[Q(t)| I_{\sigma},\, \Ev_{q,t}\right]
+\left(1- \Pp\{\Ev_{q,t}|I_{\sigma}\}\right)
\Ep \left[Q(t)| I_{\sigma},\, \NEv_{q,t}\right] \\
&\ge&  -\Pp\{\Ev_{q,t}|I_{\sigma}\}+q\left(1-
  \Pp\{\Ev_{q,t}|I_{\sigma}\}\right) \\
&=& q-(1+q)\Pp\{\Ev_{q,t}|I_{\sigma}\}\, .
\end{eqnarray}
Denote by $U_t$ the total number of spin flips in the two configurations up
to time $t$ (this is a Poisson random variable of mean $2Nt$).
Then 
\begin{eqnarray}
\Pp\{\Ev_{q,t}|I_{\sigma}\} &\le &
\Pp\{\Ev_{q,t}|I_{\sigma},U_t\le 4Nt\}+
\Pp\{U_t> 4Nt\} \\
&\le &\Pp\{\Ev_{q,t}|I_{\sigma},U_t\le 4Nt\}+
e^{-Nt/2}\, .
\end{eqnarray}
If we call $\cU$ the event $\{U_t\le 4Nt\}$, we found
\begin{eqnarray}
\tC(2t)\ge q - 2\left\{  \sum_{\sigma}\mu_{\beta}(\sigma)\,
\Pp\{\Ev_{q,t}|I_{\sigma},\,\cU\}\, +e^{-Nt/2}\right\}
\end{eqnarray}
Call $F(\sigma)\equiv \Pp\{\Ev_{q,t}|I_{\sigma},\cU\}$,
and choose $\alpha\in (1/2,1)$. By H\"older inequality
\begin{eqnarray}
 \sum_{\sigma}\mu_{\beta}(\sigma)\, F(\sigma) &=& 
\sum_{\sigma}\mu_{2\beta}(\sigma)\, 
\left(\frac{\mu_{\beta}(\sigma)}{\mu_{2\beta}(\sigma)}\right)\, F(\sigma) 
\\
&\le& \left\{\sum_{\sigma}\mu_{2\beta}(\sigma)\,
\left(\frac{\mu_{\beta}(\sigma)}{\mu_{2\beta}(\sigma)}\right)^{1/\alpha}
F(\sigma)\right\}^{\alpha}
\left\{\sum_{\sigma}\mu_{2\beta}(\sigma)\, F(\sigma)\right\}^{1-\alpha}
\label{eq:Almost}\\
& =& \frac{Z(\beta(2-\alpha^{-1}))^{\alpha}Z(2\beta)^{1-\alpha}}{Z(\beta)}
\,  \left\{\sum_{\sigma}\mu_{2\beta}(\sigma)\, F(\sigma)\right\}^{1-\alpha}\, ,
\label{eq:Last}
\end{eqnarray}
where, in passing from (\ref{eq:Almost}) to (\ref{eq:Last}), we used 
the fact that $F(\sigma)\le 1$ and the definition of $Z(\beta)$.
A moment of thought reveals that 
\begin{eqnarray}
\sum_{\sigma}\mu_{2\beta}(\sigma)\, F(\sigma) &= & 
\Pp\{\Ev_{q,t}|Q(0)=1,\cU\}
\le \frac{\Pp\{\Ev_{q,t}|U_t\le 4Nt\}}{\Pp\{Q(0)=1\}} \\
&\le& 4Nt\, \frac{\Pp\{Q(t)=q|U_t\le 4Nt\}}{\Pp\{Q(0)=1\}} 
\le 8Nt\, \frac{\Pp\{Q(t)=q\}}{\Pp\{Q(0)=1\}} \, .
\end{eqnarray}
(the last inequality follows from the fact that 
$\Pp(U_t\le 4Nt)\ge 1/2$ for $t\ge \frac{2}{N}\log 2$).
Next notice that $\Pp\{Q(t)=q\}=Z(q;\beta)/Z(\beta)^2$,
and $\Pp\{Q(0)=1\}= Z(2\beta)/Z(\beta)^2$. 
Putting the various terms together, we obtain
\begin{equation}
\tC(2 t) \ge q - 2e^{-Nt/2} - 2 (8Nt)^{1-\alpha} 
\frac{Z(\beta(2-\alpha^{-1}))^{\alpha} Z(q;\beta)^{1-\alpha} }{Z(\beta)} \, .
\end{equation}
Using Lemmas \ref{lemma:PartF} and \ref{lemma:ConstrPartF},
the product of the partition functions can be upper bounded by
$\exp[-N((1-\alpha) U(q;\beta) - \ph(\beta,\alpha)-\delta/2)]$, for any 
$\delta >0$, with high probability. With the hypothesis of the lemma
$8Nt >1$, we can thus replace $(8Nt)^{1-\alpha}$ by $8Nt$ in the inequality.
Moreover this prefactor $8N$ will be smaller than $\exp[N\delta/2]$ for
large enough $N$. We will also have  $2e^{-Nt/2}\le \delta$. 
Fixing $q$ in such a way to achieve the 
$\sup$ in Eq.~(\ref{eq:Barrier0Def}) and optimizing over $\alpha\in(1/2,1)$
completes the proof.
\endproof

We can now prove the low temperature lower bound in 
Proposition \ref{propo:TimePspin}. Consider the sum of the local correlation 
functions
\begin{eqnarray}
C(t)=\frac{1}{N}\sum_{i=1}^NC_i(t) =
\frac{1}{N}\sum_{i=1}^N[\<\sigma_i(t)\sigma_i(0)\>-\<\sigma_i(t)\>
\<\sigma_i(0)\>] \, .
\end{eqnarray}
One has $C(t) = \tC(t) - \<Q_{12}\>$, where $Q_{12}$ is the normalized
overlap of two i.i.d. configurations $\sigma^{(1)}$, and $\sigma^{(2)}$
distributed according to the Boltzmann measure
(\ref{eq:PspinBoltzmann}), (\ref{eq:PspinEnergy}). Lemma
\ref{lemma:OverlapConcentration} implies that, for any $\delta>0$, 
$C(t)\ge \tC(t)-\delta$ with high probability.
By Lemma \ref{lemma:Corr}, we have therefore
\begin{eqnarray}
C(t)\ge q_*-2\delta-t\, e^{-N[\Upsilon-\delta]}\, .
\end{eqnarray}
Fix $t_*= e^{N[\Upsilon-2\delta]}$. Then, for $N$ large enough, we
have $C(t_*)\ge q_*-3\delta$. If $f(\ve)$ is the fraction of sites $i$
such that $\tau_i(\ve)>t_*$, then $C(t_*)\le f(\ve)+\ve$,
hence $f(\ve)\ge q_*-3\delta-\ve$. The thesis follows by rescaling $\delta$.
\endproof
%
%

\end{document}